\documentclass[11pt]{article}
\pdfoutput=1 
\usepackage[DIV12]{typearea}
\usepackage[left=.9in, right=.9in]{geometry}
\usepackage[titletoc]{appendix}
\usepackage{cite}
\RequirePackage[colorlinks=true,urlcolor=blue,anchorcolor=blue,citecolor=blue,
filecolor=blue,
linkcolor=blue,menucolor=blue,linktocpage=true,pdfproducer=medialab]{hyperref}
\usepackage{amsmath,amsfonts,mathrsfs,amssymb,slashed,units,mathtools,ulem}
\usepackage{array,booktabs}
\usepackage{feynmp,graphicx,subcaption}
\graphicspath{{./arXiv/}}
\usepackage[font={small}]{caption}
\usepackage[usenames,dvipsnames]{xcolor}
\newcommand{\blue}[1]{\color{blue} #1 \color{black}}

\newcommand{\bb}{b} 

\newcommand{\s}{\sigma}

\newcommand{\ct}{c_\theta}
\newcommand{\A}{\mathcal{A}}
\newcommand{\cO}{\mathcal{O}}
\newcommand{\F}{\mathcal{F}}
\newcommand{\LL}{\mathscr{L}}
\newcommand{\U}{\mathbf{U}}
\newcommand{\T}{\mathbf{T}}
\newcommand{\V}{\mathbf{V}}
\newcommand{\Y}{\mathbf{Y}}
\newcommand{\DL}{\mathbf{D}}
\newcommand{\WL}{\mathbf{W}}
\newcommand{\vel}{\mathrm{v}}
\DeclareMathOperator{\tr}{Tr}
\renewcommand{\to}{\rightarrow}
\newcommand{\de}{\partial}
\newcommand{\nn}{\nonumber}
\newcommand{\hc}{\text{ h.c.}}
\newcommand{\email}[1]{\href{mailto:#1}{\tt #1}}

\begin{document}
\begin{titlepage}
\hypersetup{pageanchor=false}
\vspace*{-1cm}
\phantom{hep-ph/***} 
{\flushleft
{\blue{FTUAM-15-37}}
\hfill{\blue{IFT-UAM/CSIC-15-116}}}

\vskip 1.5cm
\begin{center}
{\LARGE\bf Non-linear Higgs portal to Dark Matter}\\[3mm]
\vskip .3cm
\end{center}
\vskip 0.5  cm
\begin{center}
{\large I.~Brivio}~$^{a)}$,
{\large M.B.~Gavela}~$^{a)}$,
{\large L.~Merlo}~$^{a)}$,
{\large K.~Mimasu}~$^{b)}$,
{\large J.M.~No}~$^{b)}$,
{\large R.~del~Rey}~$^{a)}$,
{\large V.~Sanz}~$^{b)}$
\\
\vskip .7cm
{\footnotesize
$^{a)}$~
Departamento de F\'isica Te\'orica and Instituto de F\'{\i}sica Te\'orica, 
IFT-UAM/CSIC,\\
Universidad Aut\'onoma de Madrid, Cantoblanco, 28049, Madrid, Spain\\
\vskip .1cm
$^{b)}$~
Department of Physics and Astronomy, University of Sussex, Brighton BN1 9QH, UK

\vskip .5cm
\begin{minipage}[l]{.9\textwidth}
\begin{center} 
\textit{E-mail:} 
\email{ilaria.brivio@uam.es},
\email{belen.gavela@uam.es},
\email{luca.merlo@uam.es},
\email{k.mimasu@sussex.ac.uk},
\email{j.m.no@sussex.ac.uk},
\email{rocio.rey@uam.es},
\email{v.sanz@sussex.ac.uk}
\end{center}
\end{minipage}
}
\end{center}

\vskip 1cm
\abstract{ 
The Higgs portal to scalar Dark Matter is considered in the context of non-linearly realised electroweak symmetry breaking. We determine the dominant interactions of gauge bosons and the physical Higgs particle $h$ to a scalar singlet dark matter candidate. Phenomenological consequences are also studied in detail, 
including the possibility of 
distinguishing this scenario from the standard Higgs portal in which the electroweak symmetry breaking is linearly realised. Two features of significant impact are: i)  the connection between the electroweak scale $v$ and the Higgs particle departs from the $(v+h)$ functional dependence, as the Higgs field is not necessarily an exact electroweak  doublet; ii) the presence of specific couplings that arise at different order in the non-linear and in the linear expansions. These facts  deeply affect the dark matter relic abundance, as well as the expected signals in direct and indirect searches and collider phenomenology, where Dark Matter production rates are enhanced with respect to the standard portal. }
\end{titlepage}
\hypersetup{pageanchor=true}
\tableofcontents

\pagebreak

\section{Introduction}

The existence of dark matter (DM) cannot be explained within the Standard Model of particle physics (SM); its discovery and that of neutrino 
oscillations constitute the first clues of {\it particle} physics beyond the SM (BSM), whose nature awaits to be revealed.
No interactions between the dark and the visible sectors have been observed\footnote{A claim for evidence of DM detection by the DAMA/LIBRA 
collaboration~\cite{Bernabei:2013xsa} has not been confirmed yet; also, some recent astrophysical analysis favouring visible-DM 
interactions~\cite{Boehm:2014vja} are open to alternative explanations~\cite{Harvey:2015hha}.} although plausibly they may exist at some 
level~\cite{Bertone:2004pz}. These putative interactions must ensure the correct DM relic abundance as well as the stability of DM on cosmological timescales.

Three types of renormalisable (marginal or relevant, i.e. dimension $d\le 4$) interactions between the SM fields and DM are possible: i) Higgs-scalar DM; ii) 
hypercharge field strength-vector DM; iii) Yukawa type couplings to fermionic DM. Being the lowest dimension  couplings of the ordinary world to DM, they are 
excellent candidates - beyond gravitational interactions - to provide the first incursions into DM, i.e. to be the experimental ``portals" into DM. In 
this paper we focus on the ``Higgs portal" to real scalar DM. 

Assuming as customary a discrete $Z_2$ symmetry~\cite{Silveira:1985rk,Veltman:1989vw} -- under which the DM singlet scalar candidate $S$ 
is odd and the SM fields are even to ensure DM stability -- the Higgs-DM portal takes the form
\begin{equation}
\lambda_{S} S^2\Phi^\dag \Phi \longrightarrow \lambda_{S} S^2 (v + h)^2 \longrightarrow \lambda_{S} S^2 (2vh + h^2)\,,
\label{SMHportal}
\end{equation} 
where $\Phi$ denotes the $SU(2)_L$ Higgs field doublet, $h$ the observed Higgs particle and $\lambda_{S}$ is the Higgs portal coupling; 
the right-hand side of the equation shows the DM-Higgs interaction in unitary gauge. The SM Higgs-DM portal in Eq.~(\ref{SMHportal}) (``standard"  portal all through this paper) has 
been extensively explored in the 
literature~\cite{Patt:2006fw,Kim:2006af,MarchRussell:2008yu,Kim:2008pp,Ahlers:2008qc,Feng:2008mu,Andreas:2008xy,Barger:2008jx,Kadastik:2009ca,
Kanemura:2010sh,Piazza:2010ye,Arina:2010an,Low:2011kp,Djouadi:2011aa,Englert:2011yb,Kamenik:2012hn,Gonderinger:2012rd,Lebedev:2012zw,Craig:2014lda}.

The nature of the Higgs particle itself also raises a quandary, though. The uncomfortable electroweak hierarchy problem -- \emph{i.e.} the surprising lightness of the Higgs particle -- remains unsolved in the absence of 
any experimental signal in favour of supersymmetry or other palliative BSM solutions in which electroweak symmetry breaking (EWSB) is linearly realised. An alternative framework is that in which EWSB is non-linearly realised (``non-linear scenario'' in short) and the lightness of the Higgs particle 
results from its being a pseudo-Goldstone boson of some global symmetry, spontaneously broken by strong dynamics at a high scale $\Lambda_s$. Much as 
the interactions of QCD pions are weighted down by the pion decay constant $f_\pi$, those of these new Goldstone bosons -- including $h$ -- will be weighted down by a 
constant $f$ such that $\Lambda_s\leq 4\pi f$~\cite{Manohar:1983md}, which may be distinct from the electroweak scale $v$ ($v\ll f$). Such an origin for a light Higgs particle 
was first proposed in the ``composite Higgs"  models in Refs.~\cite{Kaplan:1983fs,Kaplan:1983sm,Banks:1984gj,Georgi:1984ef,Georgi:1984af
}, and has been interestingly revived in recent years in view of the fine-tunings of the hierarchy problem~\cite{Contino:2003ve,Agashe:2004rs,Contino:2006qr,Gripaios:2009pe}.

An interesting characteristic of the non-linear scenario is that the low-energy physical Higgs field turns out not to be an exact electroweak doublet, and can be parametrised in the effective Lagrangian as a generic SM scalar singlet with arbitrary couplings~\cite{Feruglio:1992wf,Grinstein:2007iv,Contino:2010mh,Azatov:2012bz}. 
In other words, the typical SM dependence on $(v+h)$ in Eq.~(\ref{SMHportal}) is to be replaced by a generic polynomial $\F(h)$, implying the substitution of the standard portal  in Eq.~(\ref{SMHportal})  by the functional form
\begin{equation}
\lambda_{S} S^2 (2vh +\bb\, h^2)\,,
\label{NLHportal}
\end{equation} 
where $\bb$ is an arbitrary, model dependent constant. The $hSS$ and $hhSS$ couplings - whose relative amplitude is fixed in the standard portal -  are now decorrelated. This simple fact will be shown to have a deep impact on the estimates of the DM relic abundance, for 
which the relative strength of the DM coupling to one versus two $h$ particles plays a central role.  

A further consequence of $h$ being treated as a generic scalar singlet is that its interactions are not necessarily correlated with those of the longitudinal 
components of the $W^\pm$ and $Z$  gauge bosons, denoted by  $\pi(x)$ in the customary  $\U(x)$ matrix
\begin{equation}
\U(x)\equiv e^{i\sigma_a \pi^a(x)/v}\,. \label{Udef}
\end{equation}
While in linear BSM scenarios, $h$ and $\U(x)$ are components of the same object, i.e. the $SU(2)_L$ Higgs doublet
\begin{equation}
\label{Phi}
\Phi \equiv \frac {v+h}{\sqrt{2}} \, \U 
\left(
\begin{array}{c}
0 \\
1\\
\end{array}
\right)\,,
\end{equation}
the independence of $h$ and $\U(x)$ in the non-linear Lagrangian induces a different pattern of dominant couplings. 
Although present measurements are compatible with the SM, present Higgs data allow for sizeable departures of $h$ from being a pure Higgs doublet~\cite{Englert:1964et,Higgs:1964ia,Higgs:1964pj}. Indeed this characterisation is one of the most important quests of the LHC program, essential to unveil a putative non-linear origin of EWSB. A typical feature of the latter is the presence of relevant interactions that are expected to be further suppressed in the linear expansion~\cite{Alonso:2012jc,Alonso:2012px,Alonso:2012pz,Brivio:2013pma,Brivio:2014pfa,Gavela:2014vra,Alonso:2014wta,Hierro:2015nna} (see also Refs.~\cite{Isidori:2013cga,Buchalla:2013rka} for studies on the non-linear Higgs Lagrangian). 
It will be shown here that the bosonic couplings of $S$ also show this pattern, motivating the consideration of other interactions in addition to those 
in Eq.~(\ref{NLHportal}) above. The ensemble will lead to potential smoking guns of the nature of the EWSB mechanism and of the Higgs particle.  Distinct signals and (de)correlations in direct and collider DM searches will be discussed.

In summary, the focus of this paper is to explore the bosonic couplings of $S$ when EWSB is non-linearly realised. 
In particular, the effort will be directed to the comparison of the standard Higgs-portal encoded in Eq.~(\ref{SMHportal}) and the equivalent interactions in the ``non-linear Higgs portal''. The paper is structured as follows: in Section~\ref{Sect:Nonlinear} the purely bosonic effective Lagrangian for the non-linear Higgs portal is introduced, discussing the differences between the non-linear setup and the standard Higgs portal. In Section~\ref{Sect:Pheno} the corresponding phenomenology is worked out, analysing the DM relic abundance, direct detection and bounds from colliders. In Section ~\ref{EFT_DM_Linear} the impact of higher-dimension operators in the linear expansion is discussed and compared with the results for the non-linear portal.
In Section~\ref{Sect:Concl} we conclude.

\section{The non-linear Higgs-portal}
\label{Sect:Nonlinear}
We restrict the analysis to the purely bosonic sector, except for the fermionic Yukawa-like terms.
The relevant effective Lagrangian is derived below:  it will be shown that {\it only $v$ and the fermion and $S$ mass terms will remain as explicit scales}.

This general Lagrangian may describe the leading effects of a plethora of models, for particular values of its coefficients. In those subjacent models, aside from fermion masses, several scales may be involved explicitly and implicitly, typically:
\begin{itemize}
\item The electroweak (EW) scale $v$, at which the effective Lagrangian is defined.

\item The  Goldstone-boson scale $f$ associated to the physical Higgs $h$, whose value does not need to coincide with $v$. Arbitrary 
functions $\F(h)$ would encode the Higgs dependence as a polynomial expansion in $h$. 

\item The scale $\Lambda_s$ of the high-energy strong dynamics, with $\Lambda_s\leq 4\pi f$.

\item The  new physics scale $\Lambda_{DM}$ characteristic of the DM interactions with the visible world, that is the effective DM-Higgs portal  scale, typically corresponding to the mass of a dark mediator.

\item The mass of the scalar DM particle $m_S$. 
\end{itemize}
In the effective Lagrangian approach $v$ and the natural Goldstone boson scale $f$ are not separate parameters:
 $v$ is introduced as a fine-tuning requirement~\cite{Contino:2010rs}.  For instance it is customary to trade the $\F(h)$ polynomial dependence in powers of $h/f$ by an expansion in powers of $h/v$, with the arbitrary expansion coefficients absorbing  the $v/f$ tuning. For the heavy scales, would $\Lambda_{DM}$ coincide with $\Lambda_s$ or $f$, it would indicate a common origin for the Higgs and  the DM candidate, as it occurs in models where 
both have their origin as Goldstone bosons of the high-energy strong dynamics~\cite{Frigerio:2012uc,Marzocca:2014msa,Fonseca:2015gva}. 
Notice that, in such a scenario, the behavior of the $S$ field is expected to follow closely that of the Higgs particle: its dependence should be encoded in generic functions $\F(S)$ invariant under the $\mathbb{Z}_2$ symmetry ({\it e.g} $\cos(S/f)$).
The 
discussion will be kept here on a more general level and $\Lambda_{DM}$ will be taken as an independent scale, although assuming $f\ll\Lambda_{DM}$ in addition to plausibly $m_S\ll\Lambda_{DM}$.
 
Furthermore, only the leading terms weighted down by $\Lambda_{DM}$ and $\Lambda_{s}$ will be kept below, which in practice 
means no explicit dependence on them. Indeed, at leading order the expansion is tantamount to keeping the leading two-derivative terms of the 
electroweak chiral expansion~\cite{Appelquist:1980vg,Longhitano:1980iz,Longhitano:1980tm,Feruglio:1992wf,Appelquist:1993ka}, supplemented by the $\F(h)$ 
dependences~\cite{Alonso:2012jc,Alonso:2012px,Alonso:2012pz,Isidori:2013cga,Brivio:2013pma,Buchalla:2013rka,Brivio:2014pfa,Gavela:2014vra,
Alonso:2014wta,Hierro:2015nna,Yepes:2015zoa,Yepes:2015qwa} and the $S$ insertions: at this order the effective Lagrangian depends only  on $v$,  the fermion and $S$ mass terms, plus the operator coefficients.

The  Lagrangian can  be written as the sum of two pieces, with the second one encoding the DM interactions:
\begin{equation}
\LL=\LL_{EW}+\LL_S\,,
\label{L0}
\end{equation}
with 
\begin{equation}
\begin{split}
\LL_{EW} =
&-\dfrac{1}{4} W^a_{\mu\nu}W^{a\,\mu\nu}\F_W(h)-\dfrac{1}{4} 
B_{\mu\nu}B^{\mu\nu}\F_B(h)+\dfrac{1}{2}\de_\mu h \de^\mu h\\
&-\frac{v^2}{4}\tr(\V_\mu\V^\mu)\F_C(h)+c_T\dfrac{v^2}{4}
\tr(\T\V_\mu)\tr(\T\V^\mu)\F_T(h)-V(h)+\\
&+i\bar{Q}_L\slashed{D}Q_L+i\bar{Q}_R\slashed{D}Q_R+i\bar{L}_L\slashed{D}
L_L+i\bar{L}_R\slashed{D}L_R+\\
&-\dfrac{v}{\sqrt2}\left(\bar{Q}_L\U \Y_Q Q_R+\hc\right) 
\F_Q(h)-\dfrac{v}{\sqrt2}\left(\bar{L}_L\U \Y_L L_R+\hc\right)\F_L(h)\,,
\end{split}
\label{L0EW}
\end{equation}
where
\begin{equation}
\DL_\mu \U(x) \equiv \de_\mu \U(x) +ig\WL_{\mu}(x)\U(x) - \dfrac{ig'}{2} 
B_\mu(x) \U(x)\sigma_3 \,,
\end{equation}
with $\WL_\mu(x)\equiv W_{\mu}^a(x)\sigma_a/2$, and $W^a_\mu(x)$ and $B_\mu(x)$ denoting the SM gauge bosons. The scalar and vector chiral fields, $\T(x)$ and 
$\V(x)$, are defined as
\begin{equation}
\T(x)\equiv \U(x) \sigma_3 \U^\dagger(x)\,,\qquad\qquad
\V_\mu(x)\equiv \left(\DL_\mu \U(x)\right)\U^\dagger(x)\,,
\label{oldchiral}
\end{equation}
with transformation properties under a (global) $SU(2)_L\times SU(2)_R$ symmetry given by:
\begin{equation}
\U(x) \rightarrow L\, \U(x) R^\dagger\,,\qquad\qquad \T(x)\rightarrow 
L\,\T(x)L^\dagger\,,\qquad\qquad
\V_\mu(x)\rightarrow L\,\V_\mu(x)L^\dagger\,.
\end{equation}
After EWSB, $SU(2)_L\times SU(2)_R$ breaks down to the diagonal $SU(2)_C$, which in turn is explicitly broken by the gauged hypercharge $U(1)_Y$ and by the 
heterogeneity of the fermion masses. Equivalently, $\T(x)$ reduces to the Pauli $\sigma_3$ matrix, acting in this way as a spurion for the custodial symmetry. In Eq.~\eqref{L0EW}, the right-handed fermions have been gathered in $SU(2)_R$ quark and lepton doublets, $Q_R\equiv \{u_R,\,d_R\}$ and $L_R\equiv\{\nu_R,\, e_R \}$, while the Yukawa couplings are encoded in $\Y_{Q}\equiv {\rm diag}\{Y_U\,, Y_D\}$ and $\Y_{L}\equiv {\rm diag}\{Y_\nu\,, 
Y_\ell\}$, i.e. it assumes Higgs couplings aligned with fermion masses. This Lagrangian is akin to the SM one written in chiral notation, but for the presence of the $\F(h)$ functions and the custodial breaking $c_T$ term, which is strongly constrained by data. 

In Eq.~\ref{L0}, the DM Lagrangian $\LL_S$ at leading order in the $1/\Lambda_{DM}$ expansion reads 
\begin{equation}
\LL_S=\frac{1}{2}\de_\mu S \de^\mu S - \dfrac{m_S^2}{2} S^2 \F_{S_1}(h) - \lambda S^4 \F_{S_2}(h) + \sum_{i=1}^5 c_i\A_i(h)\,, 
\label{L0SI}
\end{equation} 
where the $\A_i$ operators form a basis:
\begin{equation}
\begin{aligned}
&\begin{rcases}
\A_1&=\tr(\V_\mu\V^\mu)S^2\F_1(h)\\
\A_2&=S^2 \square \F_{2}(h)
\hspace{3.3cm}
\end{rcases}& 
\text{Custodial Preserving}\\\\
&\begin{rcases}
\A_3&=\tr(\T\V_\mu)\tr(\T\V^\mu ) S^2 \F_{3}(h)\\
\A_4&=i\tr(\T\V_\mu) (\de^\mu S^2) \F_{4}(h)\\
\A_5&=i\tr(\T\V_\mu)S^2\de^\mu \F_{5}(h)
\hspace{1.5cm}
\end{rcases}& 
\text{Custodial Violating}
\end{aligned}
\label{scalar.op}
\end{equation} 
All $\F_{i}(h)$  functions  in  Eqs.~(\ref{L0EW}), (\ref{L0SI})  and (\ref{scalar.op}) could be generically parametrised as an expansion in powers of $h$, e.g. 
\begin{equation}
\F_{i}(h) \equiv 1 + 2\,a_i \,h/v + b_i\,h^2/v^2 + \mathcal{O}(h^3/v^3)\,.
\end{equation}
Notice, however, that no $\F(h)$ functions accompany  the Higgs, fermion and DM kinetic energies above, as they can be reabsorbed by field redefinitions without loss of generality~\cite{BGMToAppear}. Furthermore, in order to single out the impact of the DM couplings described by $\LL_S$ and to ensure a clear comparison between the chiral and the linear setups, a simplification will be adopted in what follows for the $\F_i(h)$ functions in Eq.~(\ref{L0EW}):
\begin{equation}
\F_W(h)=\F_B(h)=1\,,\qquad\qquad
\F_C(h)=(1+h/v)^2\,,\qquad\qquad
\F_Q(h)=\F_L(h)=(1+h/v)\,,
\end{equation}
while due to the strong experimental constraints on $c_T$, we safely neglect its impact. 
Finally, it is useful to rewrite $\LL_S$ as
\begin{equation}
\LL_S=\frac{1}{2}\de_\mu S \de^\mu S- \dfrac{m_S^2}{2}  S^2 
-\lambda_{S}S^2\left(2vh+\bb h^2\right)+\sum_{i=1}^5 c_i\A_i(h) +\dots 
\label{L0S}
\end{equation}
by redefining the constant parameters in an obvious way, so that the $d\le 4$ pure Higgs-DM non-linear portal takes the form announced in Eq.~(\ref{NLHportal}). The dots in Eq.~(\ref{L0S}) stand for terms with more than two $h$ bosons and/or more than two $S$ fields, which are not phenomenologically relevant in the analysis below and are henceforth discarded. 

A pertinent question is how to complete the basis including fermionic couplings. There are two possible chiral fermionic structures to consider: 
\begin{align}
&\bar{Q}_{L_i}\U Q_{R_j} S^2 \F(h)\,,\qquad\qquad \quad\,\,\,
\bar{L}_{L_i}\U L_{R_j} S^2 \F(h)\,,
\label{Fermionic1}\\
\nn\\
\begin{split}
&\bar{Q}_{L_i}\gamma_\mu Q_{L_j} \partial^\mu S^2\,\F(h)\,,\qquad \qquad
\bar{L}_{L_i}\gamma_\mu L_{L_j} \partial^\mu S^2\,\F(h)\,,\\
&\bar{Q}_{R_i}\gamma_\mu Q_{R_j} \partial^\mu S^2\,\F(h)\,,\qquad \qquad
\bar{L}_{R_i}\gamma_\mu L_{R_j} \partial^\mu S^2\,\F(h)\,,
\label{Fermionic2}
\end{split}
\end{align}
where $i,j$ are flavour indices. The equations of motion, however, allow to relate a combination of the operators in Eq.~(\ref{Fermionic1}) to the operator $\A_2$, and a combination of the operators in Eq.~(\ref{Fermionic2}) to $\A_4$. In consequence, in order to avoid redundancies, a complete basis can be defined by the ensemble of all bosonic operators in Eq.~(\ref{scalar.op}) plus those in Eqs.~(\ref{Fermionic1}) and (\ref{Fermionic2}), except for the two combinations of fermionic operators mentioned. Alternatively, the basis could be defined by all fermionic operators in Eqs.~(\ref{Fermionic1}) and (\ref{Fermionic2}) plus the bosonic ones in Eq.~(\ref{scalar.op}), excluding $\A_2$ and $\A_4$. 
An optimal choice of the basis may depend on the data considered: in this paper the focus is set on the bosonic sector only, while the effects of introducing the fermionic one deserves a comprehensive future study, where flavour effects will also be taken into account~\cite{UsFuture}

In Eq.~(\ref{L0S}),  the $c_i$'s ($i=1... 5$)  -- together with the coefficients inside $\F_i(h)$ -- parametrise the contributions of the $\A_i$ operators in the basis of Eq.~(\ref{scalar.op}).   These five effective operators describe interactions between two $S$ particles and either two $W$ bosons, one or two $Z$ or $h$ bosons, or a $Z$ and a $h$ boson 
(see the Feynman rules in Appendix \ref{Feynman_rules}), inducing interesting phenomenological signatures as shown in the next section.
 $\A_1$ and $\A_2$ are custodial invariant couplings, in the sense that they do 
not contain sources of custodial symmetry breaking other than those present in the SM (hypercharge in this case). $\A_3$, $\A_4$ and $\A_5$  provide instead new sources of 
custodial symmetry violation. Nevertheless, the contribution of $\A_4$ to the $Z$ mass  vanishes while that from $\A_5$ arises only at the two 
loop level (see Appendix~\ref{Feynman_rules}), and no significant constraint on their operator coefficient follows the $\rho$ parameter and  
EW precision data; on the other hand, these observables do receive a one-loop contribution from $\A_3$. The bound on the corresponding coefficient is estimated to be around $c_3 \sim 0.1$. 
Finally, notice that operators $\A_1$, $\A_2$ and $\A_3$ are CP-even, while $\A_4$ and $\A_5$ are CP-odd. 

In summary, the non-linear portal in Eq.~(\ref{L0S}) shows a much richer parameter space than the standard Higgs portal in Eq.~(\ref{SMHportal}).  The relationship between higher-dimension operators in the linear realisation of EWSB and the non-linear DM Higgs portal will be discussed in Section \ref{EFT_DM_Linear}.

\section{Dark Matter phenomenology}\label{Sect:Pheno}
A wide variety of experimental data constrains the DM parameter space of Higgs portal scenarios described by the Lagrangian~\eqref{L0S}. The precise measurement of the DM density today, $\Omega_{\mathrm{DM}}$, performed by Planck~\cite{Ade:2015xua} provides an upper bound on the relic abundance of $S$ particles, $\Omega_{S}$. 
Direct detection experiments set complementary limits on the strength of the DM-nucleon interactions, the current most stringent bounds coming from the Large Underground 
Xenon (LUX) experiment~\cite{Akerib:2013tjd}. Upcoming experiments like XENON1T~\cite{Aprile:2012nq,Aprile:2012zx} will further increase the sensitivity in DM direct detection. 
The couplings of DM to SM particles may be also probed at the LHC, with potential avenues including searches of invisible decay modes of the Higgs boson, 
and searches for mono-$X$ signatures, namely final states where one physical object $X$ is recoiling against missing transverse energy $\slashed{E}_T$. 

In the following we explore the rich phenomenology of non-linear Higgs portals. 
We first analyse the current constraints on the properties of DM coming from the DM relic abundance, direct detection limits from LUX and 
bounds on the invisible decay width of the Higgs boson. We then study the prospects for mono-$X$ signatures, with $X\,=$ $h$, $W^{\pm}$, $Z$, at the 13 TeV run of the LHC.
We also comment on the astrophysical signatures induced by the non-linear realisation, but defer a more detailed study of indirect detection in these models to future work.
While our phenomenological study does not intend to exhaustively explore the parameter space of non-linear Higgs portals to DM, we 
do showcase all salient features of these scenarios and confront them with the standard Higgs portal. 
A list of the observables affected by each of the new terms in the DM Lagrangian\footnote{Our analysis has some overlap with the singlet scalar case 
of \cite{Fonseca:2015gva}, which focuses on DM candidates that arise as pseudo-Goldstone bosons in specific composite Higgs models.
While it is possible to identify a correspondence between our description and theirs for the case of $\A_1$ and $\A_2$: 
$\lambda_{S} \to \bar{\lambda}$, $c_1\to d_4\, (v/f)^2$, $c_2\to a_{d_{1}}\, (v/f)^2$, in the basis of \cite{Fonseca:2015gva} there is 
no equivalent of the operators $\A_3$, $\A_4$, $\A_5$. Moreover, the $(v/f)^2$ suppression in the analysis of \cite{Fonseca:2015gva} (where 
$f=800$ GeV, $f=2.5$ TeV are considered) leads to a scan over values $\left|a_{d_{1}} \right| \times (v/f)^2< 0.1$, $\left|d_4 \right| \times (v/f)^2 < 0.1$,
corresponding to a small subset of the parameter space probed in this work.}~\eqref{L0S} is shown in Table~\ref{tab:contributions}. 

\begin{table}[t]\centering
\newcommand{\y}{\checkmark}
\begin{tabular}{lc*6{>{$}p{4.5mm}<{$}}}
\toprule
\bf Observable& &\multicolumn{6}{c}{ \bf Parameters contributing}\\
& & \bb& c_1& c_2& c_3& c_4& c_5\\
\midrule
Thermal relic density& $\Omega_Sh^2$ &  				\y& \y& \y& \y& \y&\y\\
DM-nucleon scattering in direct detection & $\s_{\mathrm{SI}}$& -	& - &  \y&  -& \y&-\\
Invisible Higgs width &$\Gamma_\text{inv}$ & 				-& -& \y&- & -&-\\
Mono-$h$ production at LHC& $\s(pp\to hSS)$& 				\y& -& \y& - & \y&\y\\
Mono-$Z$ production at LHC& $\s(pp\to ZSS)$ &				-& \y& \y& \y& \y&\y\\	
Mono-$W$ production at LHC& $\s(pp\to W^+SS)$ & 			-& \y& \y&- & \y&-\\
\bottomrule
\end{tabular}
 \caption{Non-linear Higgs portal parameters affecting each of the observables considered. The standard Higgs-DM portal $\bb=1$ and all $c_i$=0.}\label{tab:contributions}
\end{table}

The non-linear DM-Higgs portal from Eq.~\eqref{L0S} is implemented in {\tt FeynRules}~\cite{Alloul:2013bka} and interfaced to {\tt MicrOMEGAs}~\cite{Belanger:2014vza} and {\tt MadGraph5$\_$aMC@NLO}~\cite{Alwall:2014hca} to compute the relevant observables. For the analysis of mono-$X$ 
signatures at the LHC, we use in addition the 1-loop {\tt FeynRules/NLOCT} implementation of gluon-initiated mono-$X$ signatures via an s-channel mediator 
from \cite{Mattelaer:2015haa}, in order to capture the full momentum dependence in the production of mono-$X$ signatures via gluon fusion. In all cases, the standard  portal 
corresponds to the choice $\bb=1,\, c_i=0$, and we compare it with different non-linear portal setups in which one of the parameters of the set \{$b,\, c_i$\} is varied at a time. 
This approach ensures a clear and conservative phenomenological comparison between the standard and the non-linear portal scenarios, allowing to single out the physical impact 
of each effective operator. 

Finally, a comment on the range of validity of the analysis is in order: while the couplings studied do not depend on the actual value of $\Lambda_{DM}$, our results
should only be taken as indicative when involving scales ($m_S$ or $p_T$) above $1$ TeV, as the heavy scale $\Lambda_{DM}$ cannot plausibly be much larger while 
still having an impact on the present and foreseen experimental sensitivities.

\begin{figure}[t!]\centering
\includegraphics[width=11cm]{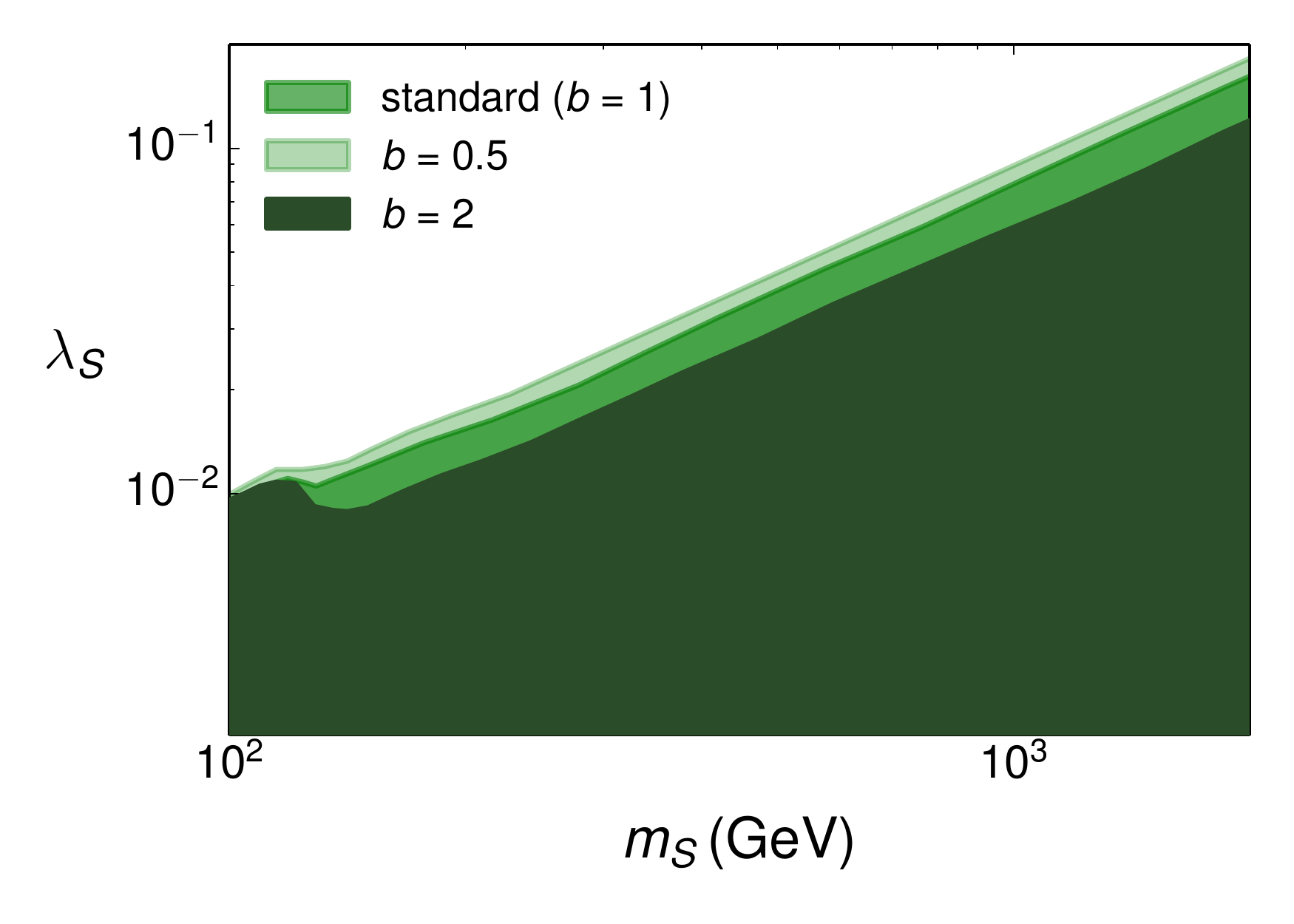}
\caption{Regions excluded by the condition $\Omega_S h^2 \leq 0.12$ for DM masses $m_S\geq \unit[100]{GeV}$. 
The medium green region corresponds to the standard Higgs portal case $\bb=1$, while the light/dark green regions (superimposed) 
correspond respectively to $\bb=0.5$ and $\bb=2$.}\label{plot_omegab}
\end{figure}

\subsection{Dark Matter relic density}
Assuming that the singlet scalar particle $S$ is a thermal relic, its abundance $\Omega_S$ today is determined by the thermally averaged
annihilation cross section into SM particles in the early Universe $(\s \vel)_\text{ann}=\s(SS\to XX)\,\vel$. For non-relativistic relics, this cross section can be expanded as
\begin{equation}
(\sigma \vel)_\text{ann} =\alpha_s+ \alpha_p\, \vel^2
\end{equation}
where $\alpha_s$ is the (unsuppressed) $s$-wave contribution, and the next order in the expansion, $\alpha_p$, corresponds to the $p$-wave contribution. Noticing 
that $\langle \vel \rangle^2 = 6/x_F$, with $x_F$ given by the freeze-out temperature as $x_F = m_S/T_F \simeq 20$, the relic density is determined by
\begin{equation}
\Omega_{S} h^2\simeq\frac{2.09\times10^{8}\,{\rm 
GeV}^{-1}}{M_P\sqrt{g_{*s}(x_F)}(\alpha_s/x_F+3 \,\alpha_p/x^2_F)} \ ,
\end{equation}
with $M_P$ being the Planck mass and $g_{*s} (x_F)$ the number of relativistic degrees of freedom at a temperature $T_F$.
The $s$-wave contributions to the DM annihilation cross-section for the different channels (the corresponding tree-level Feynman diagrams are shown in 
Appendix \ref{Relic_Abundance}) are given by 
\begin{align}
\label{SSff}
\alpha_s (S \, S \to f \bar{f}) &=\frac{6\lambda_{S}^2}{\pi  m_S^2 }\,
\frac{r_f^2(1-r_f^2)^{3/2}}{ \left(r^2-4\right)^2} \left[\left(1+\frac{4c_2a_2}{r_v^2}\right)^2+\frac{c_4^2}{r_v^4}\frac{(r^2-4)^2}{1-r_f^2}\right]\,,
\\
\alpha_s(S \, S \to h \, h) &= \frac{\lambda_{S}^2}{8 \pi m_S^2} \,  \frac{\sqrt{1-r^2}\,K_{h_{0}}^2}{(r^4-6 r^2+8)^2}\Bigg[1+
\frac{4c_2a_2}{K_{h_{0}}}\frac{r^2}{r_v^2}\Big(4r_v^2(r^2-4)-3(r^2-2)+\nn\\
&\quad +2c_2a_2r^2(r^2-4)+\frac{b_2}{a_2}\frac{r^4-6r^2+8}{r^2}\Big)\Bigg]^2\,,\\
\alpha_s (S\, S\to Z\, Z) &= \frac{\lambda_{S}^2}{8\pi m_S^2}\frac{\sqrt{1-r_Z^2}}{(r^2-4)^2} K_{Z_{0}}\left[1+\frac{4c_2a_2}{r_v^2}+(c_1+2c_3)\frac{r^2-4}{r_v^2}\right]^2\,,\\
\label{SSWW}
\alpha_s (S\, S\to W^+\, W^-) &=  \frac{\lambda_{S}^2}{4\pi m_S^2}\frac{\sqrt{1-r_W^2}}{(r^2-4)^2}K_{W_{0}}\left[1+\frac{4c_2a_2}{r_v^2}+c_1\frac{r^2-4}{r_v^2}\right]^2\,,\\
\alpha_s (S\, S\to Z\, h) &= \frac{\lambda_{S}^2}{512\pi m_S^2}\frac{\left[(r^2+r_Z^2-4)^2-4r^2r_Z^2\right]^{3/2}}{r_v^4}(2c_4+c_5a_5)^2\,\,,
\end{align}
with $r=m_h/m_S$, $r_f=m_f/m_S$, $r_{Z,W}=m_{Z,W}/m_S$, $r_v=\sqrt{\lambda_{S}}\, v/m_S$ and $K_{h_{0}}$, $K_{Z_{0}}$, $K_{W_{0}}$ defined as 
\begin{align}
K_{h_{0}}&= (\bb-3) r^4-6 (\bb-1) r^2+8 \bb+8 \left(r^2-4\right) r_v^2\,,\\
K_{Z_{0}} &=4(1-r_Z^2)+3r_Z^4\,,\\
K_{W_{0}} &=4(1-r_W^2)+3r_W^4\,.
\end{align}
Each annihilation channel contains, in general, new non-linear pieces in addition to the standard contributions, including the decorrelations from $\bb$ in the $SS\to hh$ channel. The sole exception to this behaviour is 
the annihilation channel $S S \to Z h$, which receives contributions from the CP-violating operators $\A_{4,5}$ and is absent in the standard case, inducing an $s$-wave 
leading term proportional to $c^2_{4,5}$. 

In the following we discuss how non-linear contributions change the predictions for the Higgs portal. In a conservative approach, we require the abundance of $S$ particles 
today not to exceed the total DM density measured by Planck~\cite{Ade:2015xua}, imposing $\Omega_{S} h^2 \leq \Omega_{\mathrm{DM}} h^2 \simeq 0.12$ but not requiring 
$S$ to account for the entire DM relic abundance\footnote{This constitutes another important difference with the analysis of Ref.~\cite{Fonseca:2015gva}, which requires the scalar 
singlet $S$ to constitute all the DM. Although a direct comparison of our results with those of Ref.~\cite{Fonseca:2015gva} is then difficult due to the different analysis methodology, we can state that our conclusions are compatible with theirs.}. 
Let us start by discussing the non-linear mismatch between the terms which are linear and quadratic in Higgs fields, parametrised by the coefficient
$\bb$ in Eq.~\eqref{L0S}. Values $\bb \neq 1$ modify the relative strength of the $SShh$ and $SSh$ couplings {\it w.r.t.} the standard Higgs portal. 
This mismatch can be observed in the region $m_S> m_h$, where the annihilation into two Higgs bosons is important. As shown in Figure~\ref{plot_omegab}, 
for $\bb > 1$ the annihilation cross section into Higgses increases significantly, thus enlarging the allowed region of parameter space for the non-linear portal. 

Consider now the impact of the non-linear $\A_i$ operators on $\s_\text{ann}$.  
Operators $\A_{1-5}$ affect DM annihilations into gauge bosons, Higgses and $b$-quarks, as shown in Appendix \ref{Relic_Abundance}.
This modifies the relic density $\Omega_{S}$ both for large and small values of $m_S$.
\begin{figure}[t!]
\hspace*{-1.2cm}
\includegraphics[height=6.5cm]{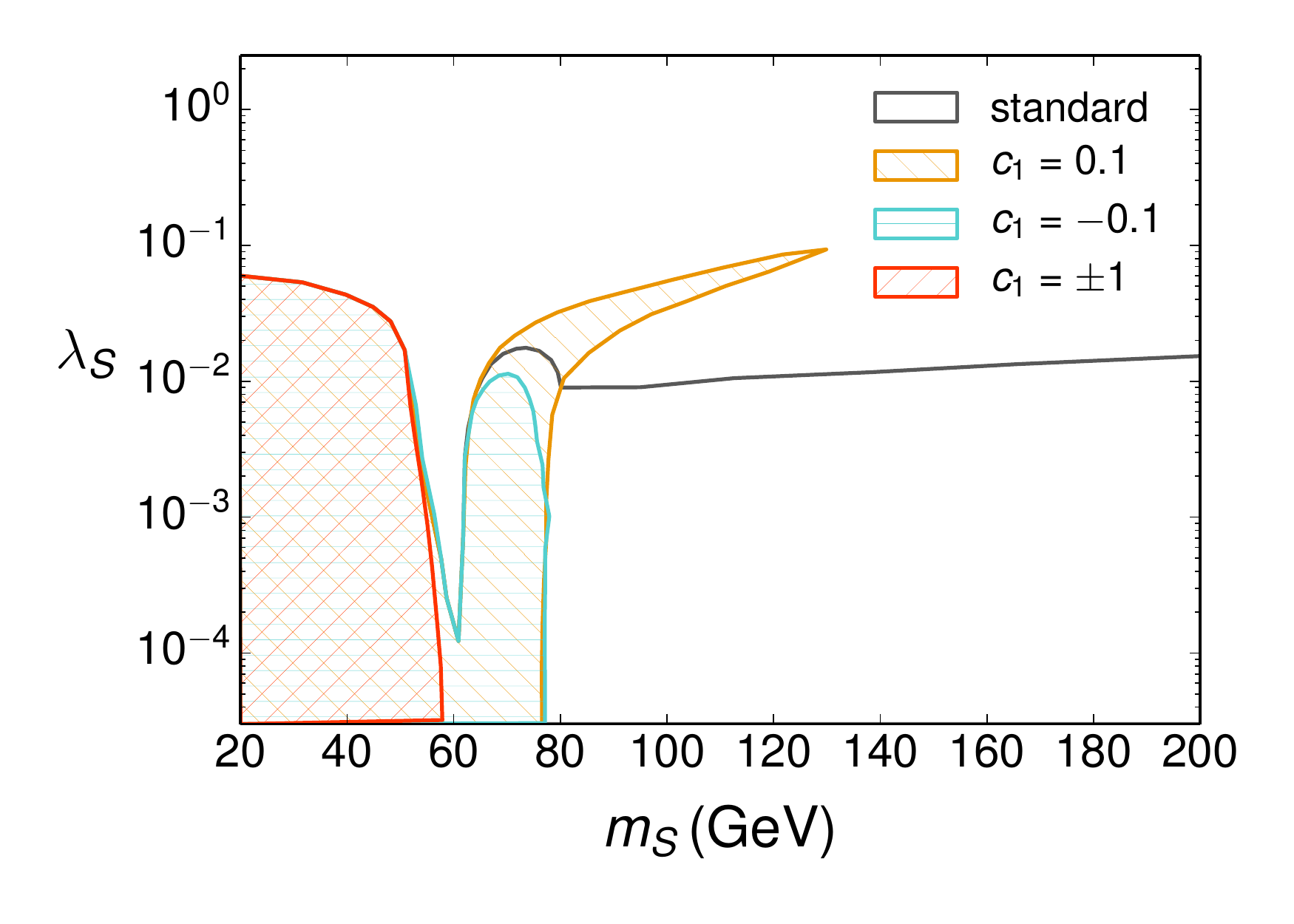}
\hspace*{-4mm}
\includegraphics[height=6.5cm]{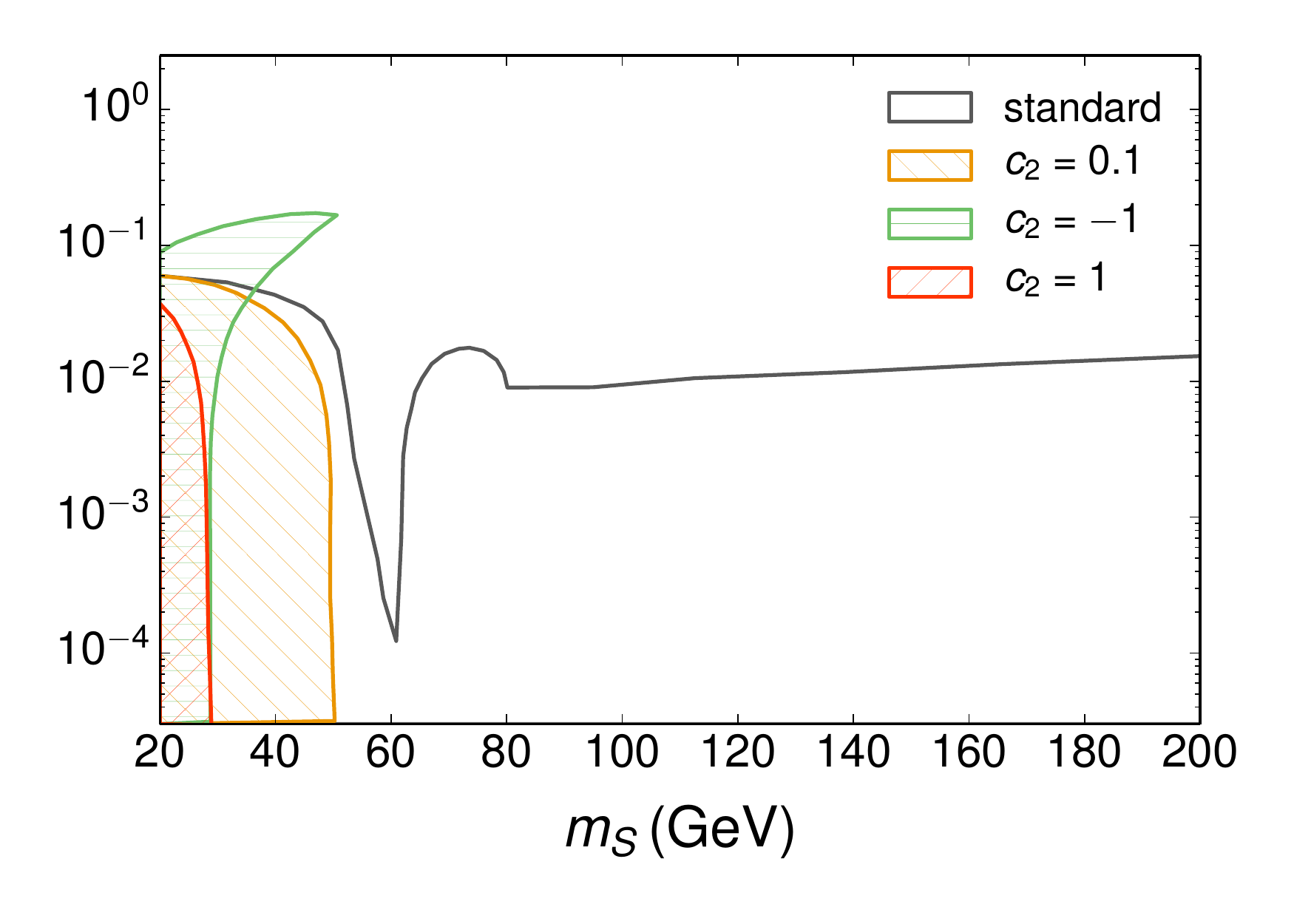}
\caption{Regions in the ($m_S$, $\lambda_S$) plane excluded by the constraint $\Omega_Sh^2\leq 0.12$ from Planck~\cite{Ade:2015xua}, in presence of non-linear operators
$\A_1$ (Left) and $\A_2$ (Right) with $c_i\neq0$. The region below the black line is excluded for the standard Higgs portal. 
Left: excluded regions for $c_1 = 0.1$ (yellow), $c_1 = -0.1$ (light blue), $\left|c_1\right|= 1$ (red). Right: excluded regions for $c_2 = 0.1$ (yellow), $c_2 = 1$ (red), $c_2 = -1$ (green).}\label{plots_omega}
\end{figure}
To illustrate these new effects, we compare in Figure~\ref{plots_omega} the parameter space excluded for the standard Higgs 
portal (our results for the standard Higgs portal scenario are in agreement with those of Refs.~\cite{Cline:2013gha,Duerr:2015mva,Duerr:2015aka,Han:2015hda})
and in the presence of the custodially-preserving and CP-even operators $\A_1$ and $\A_2$, with $c_1$, $c_2$ in the 
range $[-1,1]$. It shows the  drastic increase resulting in the parameter space for DM masses larger than tens of GeV, as compared with the allowed region for the standard portal above the black curve.
For simplicity, in this figure the dependence on the Higgs field is fixed to be $\F_1(h)=\F_2(h)=(1+h/v)^2$, 
corresponding to $a_1=b_1=a_2=b_2=1$; we have checked that varying these values does not change noticeably the impact on the dark matter relic density 
$\Omega_S h^2$, as expected~\footnote{$a_1$ ($b_1$) parametrises vertices $SSVVh$ ($SSVVhh$), with $V= Z,\,W^\pm$, whose tree-level contribution to the DM annihilation 
cross section  is very much suppressed due to phase space considerations; a variation of $a_2$ can be reabsorbed in the normalisation of $c_2$; finally, $b_2$ enters the $SS\to hh$ cross-section 
for masses $m_S>m_h$, but its effect is only significant for unrealistically large values of $b_2$.}.   

\vspace{2mm}

In the presence of $\A_1$, DM can directly interact with SM gauge bosons via the vertices $SSZZ$ and $SSW^+W^-$. 
 The new interactions induced by $\A_1$ do not modify the
allowed parameter space for $m_S \lesssim 65$ GeV, where DM annihilates dominantly into $b\bar{b}$, while they have a strong impact 
on the DM annihilation process into two gauge bosons, 
which becomes important as $m_S$ grows, as shown in Figure~\ref{plots_omega} (Left).  
For negative values of $c_1$, the positive interference with the linear amplitude (see the Feynman rules in 
Appendix \ref{Feynman_rules}) increases the total annihilation cross-section everywhere and some of the points ruled out in the standard Higgs portal scenario become 
viable. On the other hand, if $c_1>0$ the interference is destructive and spurious cancellations may happen in regions of the parameter space that are allowed for 
standard Higgs portals, but become now excluded. As an example, the yellow ``branch'' structure in Figure~\ref{plots_omega} (Left) for $60$ GeV $\lesssim 
m_S \lesssim 130$ GeV is traversed by a curve on which $\alpha_s(SS\to VV)=0$ for $V= Z,\,W^\pm$.  
\vspace{2mm}

The impact of the operator $\A_2$, shown in Figure \ref{plots_omega} (Right), can be understood in an analogous way: the 
coefficient $c_2$ enters the couplings $SShh$ and $SSh$, with the double effect of boosting the $SS\to hh$ 
process for $c_2>0$ and generating local cancellations when $c_2<0$ on one side, and also altering the annihilation $SS \to b\bar{b}$ through an $s-$channel 
Higgs, which significantly affects the annihilation cross section below $m_S \simeq m_h/2$.

\vspace{2mm}
The operator $\A_3$ has a similar phenomenology to that of $\A_1$, although restricted exclusively to DM annihilation into $Z$ bosons (at tree level). However, the presence of 
$\A_3$ is tightly constrained by EW precision data (see the discussion at the end of Section \ref{Sect:Nonlinear}). As the present bound on $c_3$ is already below the foreseen experimental sensitivities we 
will not further analyze it separately.

\subsection{Direct detection of Dark Matter}
DM interactions with nucleons are probed at direct detection experiments, which provide upper limits on the spin-independent and spin-dependent 
cross-sections. The scalar $S$ interacts with fermions via the Higgs and, in the non-linear case, via $W^\pm$ and $Z$ exchange. The most important constraints in our 
scenario come from the stronger spin-independent limits, which give an upper bound on the cross section $\s_{\mathrm{SI}}$ for scattering of $S$ on nucleons. 
$S$ may not be the only DM particle, but a member of a new DM sector, and in this case $\Omega_{S} < \Omega_{\mathrm{DM}}$. When translating bounds on direct 
detection cross-section one can account for this fact by the following rescaling
\begin{equation}
\label{DD_bound}
 \s_{\mathrm{SI}}(S\,N\to S\,N)\, \frac{\Omega_S}{\Omega_{\mathrm{DM}}}\leq 
\s_\text{exp}^{lim}\,,
\end{equation}
where $\s_\text{exp}^{lim}$ is the experimental upper limit on the DM-nucleon scattering cross-section. Here we 
consider the current most stringent 95\% Confidence Level (C.L.) experimental limits by LUX~\cite{Akerib:2013tjd}, as well as the 95\% C.L. projected sensitivity of 
XENON1T~\cite{Aprile:2012zx}.

\begin{figure}[t!]\centering
\includegraphics[width=0.7\textwidth]{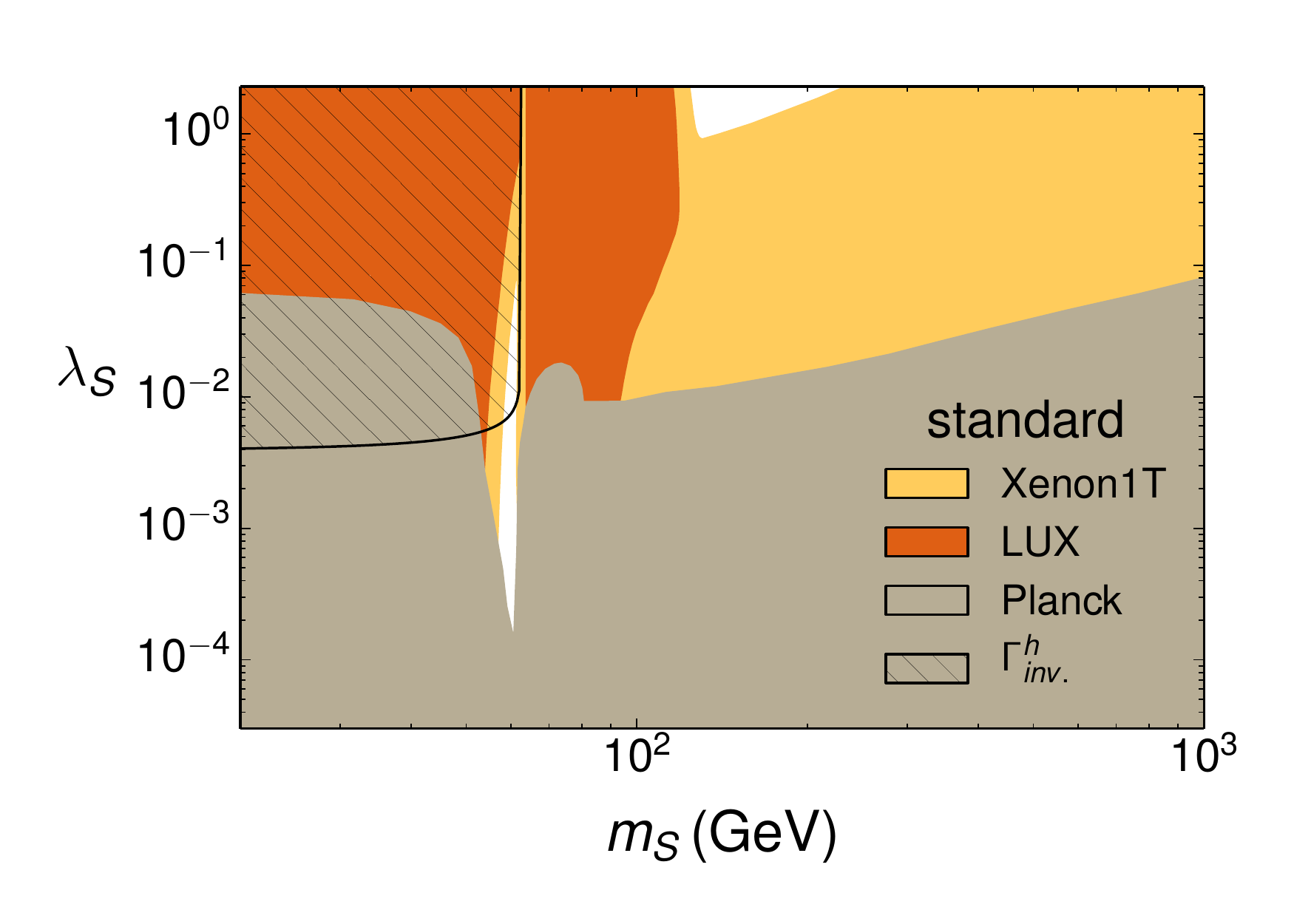}
\caption{Standard Higgs portal (corresponding to the case $c_i\equiv0,\,\bb=1$) in the ($m_S,\,\lambda_S$) plane, for masses $m_S$ up to 1 TeV. 
The grey region is excluded by current bounds from Planck~\cite{Ade:2015xua}.
The orange region is excluded by LUX~\cite{Akerib:2013tjd}, while the yellow area is currently allowed but within the reach of XENON1T~\cite{Aprile:2012zx}. The black-hatched region 
represents the region excluded from the invisible Higgs width data (see Section \ref{Section_HiggsWidth}).}\label{plots_summarylin} 
\end{figure}

\begin{figure}[t!]
\begin{subfigure}{.495\textwidth}\centering
\hspace*{-1cm}
\includegraphics[height=6.6cm]{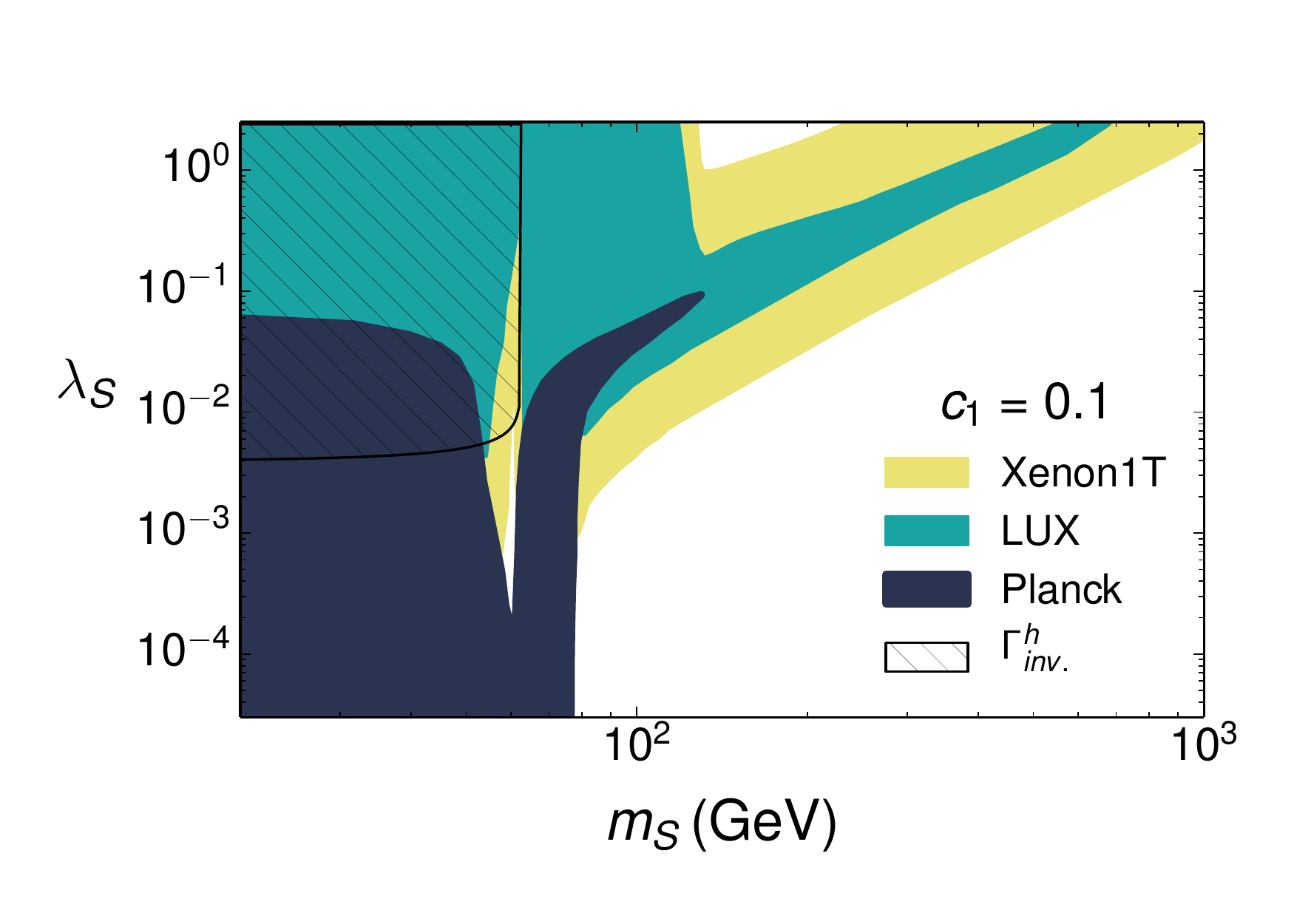}
\caption{$c_1=0.1$}\label{plot_summaryc101}
\end{subfigure}
\begin{subfigure}{.495\textwidth}\centering
\includegraphics[trim = 1.3cm 0cm 0cm 0cm, clip=true, totalheight=6.6cm]{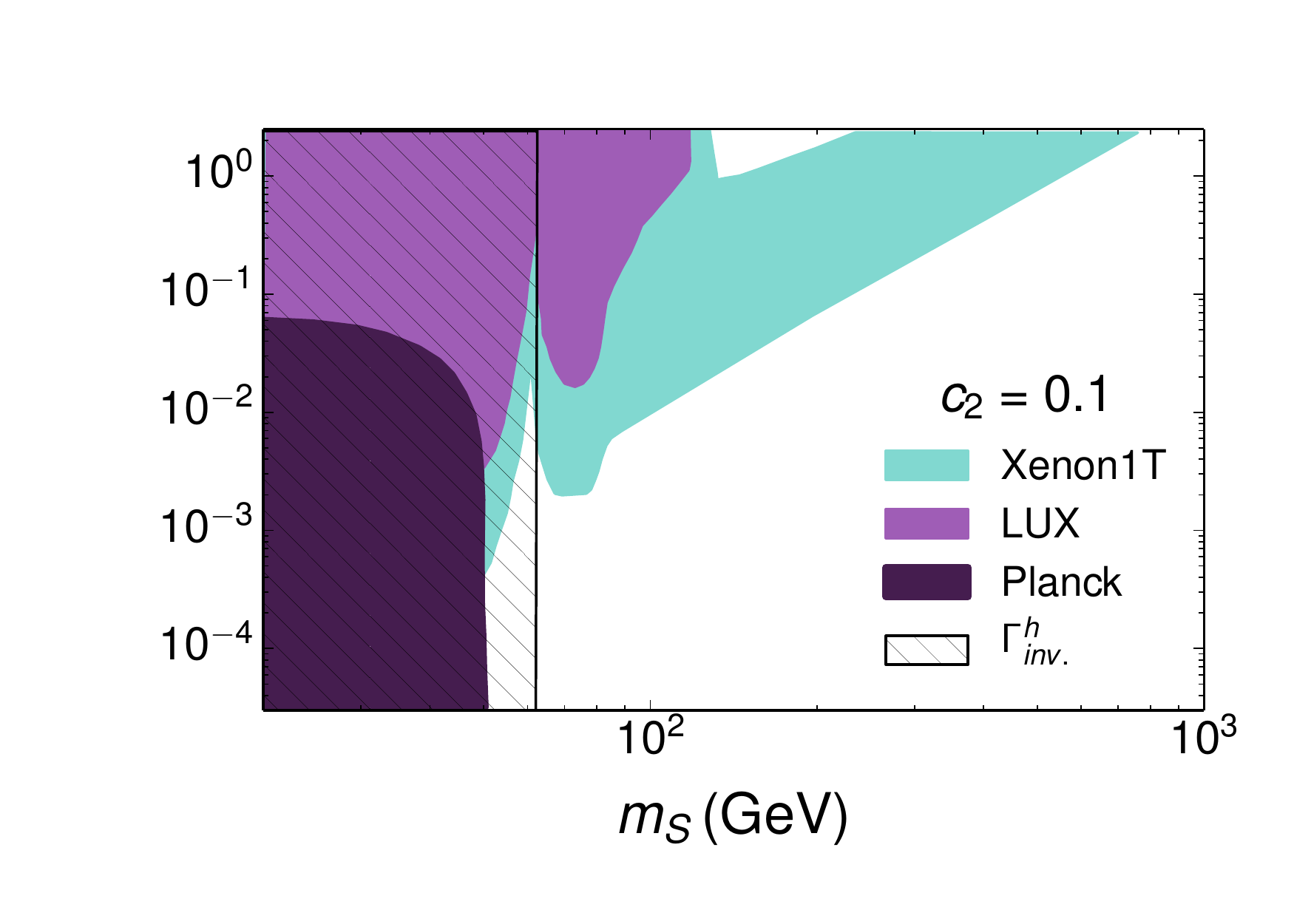}
\caption{$c_2=0.1$}\label{plot_summaryc201}
\end{subfigure}
\caption{Non-linear Higgs portals in the ($m_S$, $\lambda_S$) plane, considering the non-linear operators $\A_1$ (Left) and $\A_2$ (Right) 
with $\F_i(h)=(1+h/v)^2$ and $c_i=0.1$. The darkest region is excluded by current bounds from Planck, the green/purple one is excluded by LUX, while 
the area in yellow/light blue is within the projected reach of XENON1T.  The black hatched region represents the bound 
from the invisible Higgs width (see Section \ref{Section_HiggsWidth}).}\label{plots_summarynonlin} 
\end{figure}

The white areas in Figure~\ref{plots_summarylin} and \ref{plots_summarynonlin} summarise the DM parameter space allowed by Planck data and lying below the XENON1T direct detection sensitivity reach, for the standard and non-linear portals respectively. Specifically, the current and 
projected direct detection exclusion regions in the plane $(m_S,\,\lambda_S)$ obtained with 
\texttt{MicrOMEGAs} are shown in Figure~\ref{plots_summarylin} for the standard Higgs portal scenario, and in Figure~\ref{plots_summarynonlin} 
in the presence of the non-linear operators $\A_1$ or $\A_2$ with a coefficient $c_i=0.1$, fixing for simplicity $\F_1(h)=\F_2(h)=(1+h/v)^2$ (see footnote 4). The following discussion 
will be restricted to these two cases, that exemplify quite exhaustively the main features introduced by non-linearity. For further scenarios corresponding to different 
choices of the coefficients $c_1,\,c_2$ in the range $[-1,1]$ we defer the reader to Appendix~\ref{c12_extras}.
We stress that, while neither $\A_1$ nor $\A_2$ affect the $S$-nucleon 
scattering cross-section to first approximation ($\A_1$ gives $SSZZ$ and $SSW^{+}W^{-}$ vertices which do not enter the scattering at tree level, while the contribution 
of $\A_2$ is proportional to the transferred momentum, and thus highly suppressed at such low energies), the impact of these two operators on the relic abundance 
$\Omega_S$ affects the direct detection exclusion regions, as shown in Figure~\ref{plots_summarynonlin}.
It is also worth noting that, despite providing an independent and complementary bound to that from the Planck Satellite, the direct detection results share 
some features with those obtained imposing the constraint by Planck. As discussed in the previous section, the allowed portion of parameter space is 
generically enlarged for either $c_1<0$ or $c_2>0$ compared to the standard case (see Figure~\ref{plot_summaryc201}), while for $c_1>0$ or $c_2<0$ the exclusion region may 
occasionally stretch further into an area that is allowed in the standard setup, as in Figure~\ref{plot_summaryc101}.

Let us also comment on the impact of the operator $\A_4$ on DM-nucleon scattering: as shown in Appendix~\ref{Feynman_rules}, this operator 
induces an effective vertex $SSZ$ that allows a diagram for the $qS\to qS$ process with a $Z$ boson mediating in $t$-channel. However, the corresponding contribution to the squared amplitude is proportional to the Mandelstam variable, $t$:
\begin{equation}
 \left|A(q S\to q S)\right|^2 \sim c_4^2\,\frac{g^4}{(c_{\theta_W})^4}\,\frac{m_q^2}{m_Z^4}\,t \,
\end{equation}
with $c_{\theta_W}$ denoting the cosine of the Weinberg angle. This contribution then vanishes in the limit of zero transferred 
momentum $t\to 0$. As a result, the coefficient $c_4$ is not bounded by direct detection experiments, a conclusion that we have independently verified using {\tt MadDM}~\cite{Backovic:2015cra}.

\subsection{Invisible Higgs decay width}
\label{Section_HiggsWidth}
A very powerful probe of Higgs portal DM in the mass region $m_S<m_h/2$ is given by searches for an invisible decay 
width of the Higgs boson at the LHC. The decay $h\to SS$ is open for $m_S<m_h/2$, and contributes to the Higgs invisible width $\Gamma_\text{inv}$ as
\begin{equation}
\Gamma_\text{inv}= \frac{\lambda_{S}^2v^2}{2\pi 
m_h}\sqrt{1-\frac{4m_S^2}{m_h^2}}\left(1+\frac{c_2 a_2 
m_h^2}{\lambda_{S}v^2}\right)^2\,.
\label{H_Invisible_Width}
\end{equation}
As is clear from Eq.~(\ref{H_Invisible_Width}), the presence of $\A_2$ gives a further contribution to $\Gamma_\text{inv}$ \textit{w.r.t.} the standard Higgs portal,
such that, if $c_2 a_2 \neq 0$, then $\Gamma_\text{inv} > 0$ even for $\lambda_{S} \to 0$. Current experimental searches by ATLAS \cite{Aad:2015txa,Aad:2015pla} 
and CMS \cite{Chatrchyan:2014tja} constrain the $h\to$ \textsl{invisible} branching fraction, with the strongest limit requiring~\cite{Aad:2015pla}
\begin{equation}
\text{BR}_\text{inv} = 
\frac{\Gamma_\text{inv}}{\Gamma_\text{inv}+\Gamma_\text{SM}} < 0.23 \quad\,\, 
(95\%\, \text{CL})
\label{H_Invisible_Width_ATLAS}
\end{equation}
where the SM width is $\Gamma_\text{SM}\simeq\unit[4]{MeV}$. Conveniently setting the parameter $a_2 = 1$ (as it can always 
be reabsorbed in the normalization of $c_2$), we present the exclusion region obtained from Eqs. (\ref{H_Invisible_Width}) and (\ref{H_Invisible_Width_ATLAS})
as a black hatched area in Figures \ref{plots_summarylin} and \ref{plot_summaryc101} for $c_2=0$, and Figure \ref{plot_summaryc201} for $c_2=0.1$. For Figure~\ref{plot_summaryc101} the limit coincides 
with the one derived for the standard Higgs portal plotted also in Figure~\ref{plots_summarylin} (see e.g. \cite{Cline:2013gha,Duerr:2015mva,Duerr:2015aka,Han:2015hda}), while Figure \ref{plot_summaryc201} illustrates the effect of $c_2 \neq 0$: even for 
small values of this coefficient, the bound becomes very stringent, with practically all the region $m_S<m_h/2$ being excluded.

\begin{figure}[t!]\centering
\includegraphics[width=.65\textwidth]{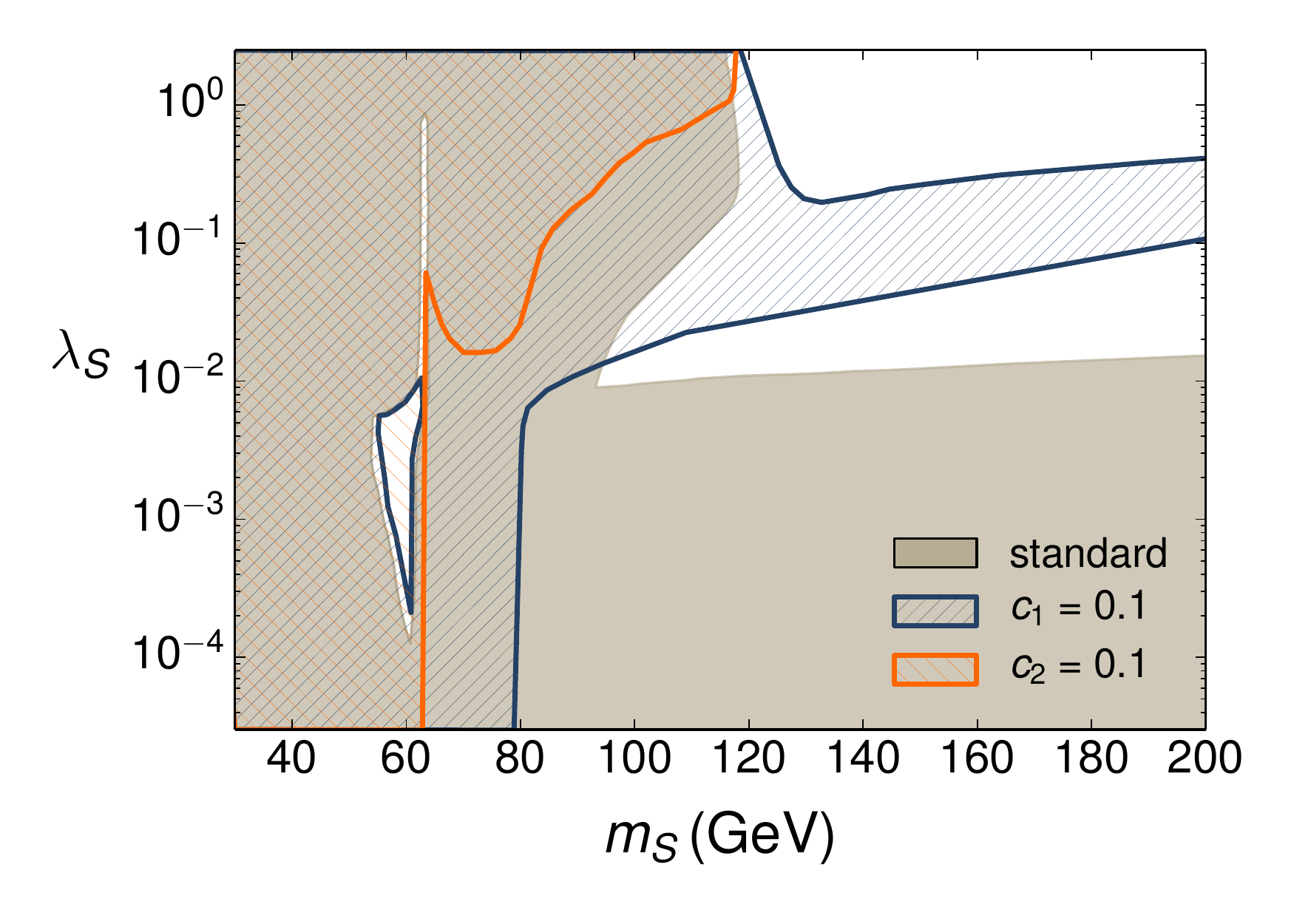}
\caption{Current excluded region in the ($m_S$, $\lambda_S$) plane for the standard Higgs portal (grey) versus the non-linear one for $c_1=0.1$ (blue) and $c_2=0.1$ (orange), from DM relic abundance, direct detection 
and invisible decay width of the Higgs.}\label{plots_omega_current}
\end{figure}

\vspace{2mm}

It is important to stress that, even though the non-linear operator $\A_4$ generates a $SSZ$ vertex, the $Z$ invisible width is not affected by it. The would-be contribution from $\A_4$ is CP-odd and also vanishes whenever the $Z$ is on-shell.

\vspace{2mm}

The impact of non-linear contributions on the parameter space of Higgs portals, combining the information from the DM relic density, direct detection experiments and 
searches for invisible decay modes of the Higgs boson is exemplified in Figure \ref{plots_omega_current}, which shows the comparison between the combined excluded region 
for the standard Higgs portal (grey region) and the combined excluded regions in the presence of $\A_1$ with $c_1 = 0.1$ (hatched-blue region) and 
in the presence of $\A_2$ with $c_2 = 0.1$ (hatched-orange region).

\subsection{Dark Matter at the LHC: Mono-X searches}

As already highlighted in the previous section, the LHC (and collider experiments in general) constitutes a natural place to search for DM interactions with the SM, in particular 
if such interactions involve the EW sector of the theory. LHC probes of DM provide an independent test of the results from low-energy and astrophysical 
experiments, while being able to directly explore a new energy regime.

\vspace{2mm}

A key probe of DM production at colliders are ``mono-$X$'' signatures, {\it i.e.}~the associated production of DM particles with a visible object $X$, which is seen 
to recoil against a large amount of missing transverse energy $\slashed{E}_T$. These signatures are in principle sensitive to relatively large DM masses, but for the standard Higgs 
portal scenario the relevant cross sections at the LHC drop very quickly for $m_S>m_h/2$, making it challenging to extract information on the DM properties
from these searches (see {\it e.g.} \cite{Craig:2014lda}). As we show below, the presence of non-linear Higgs portal interactions $\A_{1-5}$ has a dramatic impact on the LHC potential for probing such 
mono-$X$ signatures. 

\begin{figure}[t!]\centering
\begin{subfigure}{\textwidth}\centering
\input{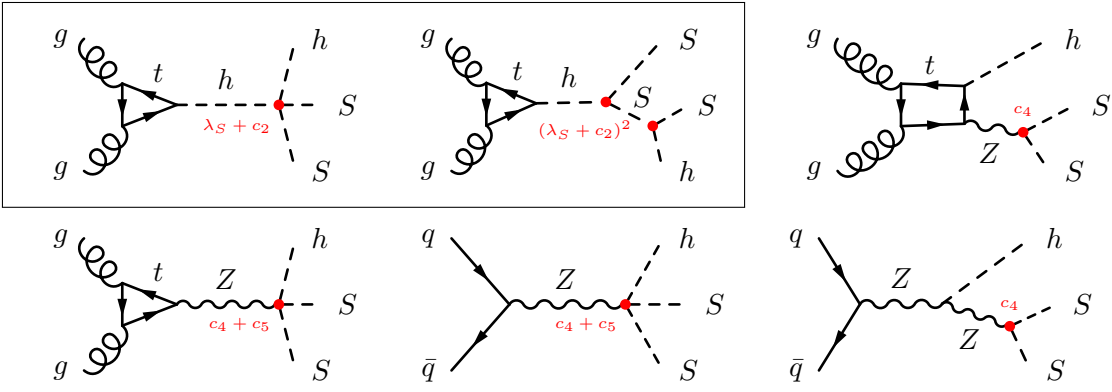}
\end{subfigure}
\caption{Sample of the main Feynman diagrams contributing to mono-$h$ production. 
In the standard Higgs case only those inside the frame are present: the process is entirely $gg$-initiated, with contributions proportional to $\lambda_S$ and to $\lambda_S^2$.
In the non-linear scenario all the diagrams contribute: both $gg$- and $\bar{q}q$-initiated processes are included. The proportionality of each diagram to the non-linear parameters is indicated in the figure (overall factors and numerical coefficients are not specified).
}\label{diagram_monoHprod}
\end{figure}

\vspace{2mm}

We focus our analysis on mono-$h$, mono-$Z$ and mono-$W$ signatures at the LHC, and present a detailed comparison of the standard and non-linear Higgs portal DM scenarios in this context. We stress that for the case of mono-$h,Z$ signatures, both $\bar{q} q$ and gluon ($g g$) -initiated processes are possible. The latter 
are characterised by loop-induced DM production processes, which we compute using the {\tt FeynRules/NLOCT} framework~\cite{Degrande:2014vpa} interfaced to {\tt MadGraph5$\_$aMC@NLO} 
and {\tt MadLoop}~\cite{Hirschi:2011pa,Hirschi:2015iia},
to ensure that the momentum dependence of the loop is accurately described. This particular aspect is crucial for a meaningful comparison between the standard and non-linear Higgs portal scenarios.

\subsubsection{Mono-\texorpdfstring{$h$}{h} signatures}\label{Section_monoH}
Mono-Higgs searches~\cite{Petrov:2013nia,Carpenter:2013xra,Berlin:2014cfa,No:2015xqa} have been proposed recently 
as a probe of the DM interactions with the SM, particularly in the context of Higgs portal scenarios. This proposal has led 
the ATLAS experiment to perform a search for mono-$h$ signatures in the $\slashed{E}_T + \gamma \gamma$~\cite{Aad:2015yga} and $\slashed{E}_T+b\bar{b}$~\cite{Aad:2015dva} final states with 20.3 fb$^{-1}$ of LHC 8 TeV data. While the latter channel is not conclusive for the case of scalar Dark Matter, the former yields a 95\% C.L. limit on the mono-$h$ fiducial cross section $\s^{\gamma\gamma}_{\text{mono-$h$}}\leq \unit[0.7]{fb}$ (with $h \to \gamma \gamma$) 
after the selection $\slashed{E}_T > 90$ GeV. 

\vspace{2mm}

For the standard Higgs portal, mono-$h$ processes are $g g$-initiated
and the amplitude receives contributions from Feynman diagrams scaling as $\sim \lambda_S$ and $\sim \lambda^2_S$, as depicted in Figure \ref{diagram_monoHprod} (within the frame), 
the latter providing a significant enhancement in the cross section when $\lambda_S \sim 1$. We note however that for $\lambda_S = 1$, satisfying the direct detection bound from LUX 
requires $m_S > 127$ GeV (see Figure \ref{plots_summarylin}), and for that range of masses the mono-$h$ cross section gets suppressed due to the intermediate off-shell Higgs state and the steep fall of the gluon PDF at high $\sqrt{\hat{s}}$. Moreover, 
limits from the invisible decay width of the Higgs require $\lambda_S \lesssim 0.007$ for $m_S < m_h/2$ in this scenario (see Figure \ref{plots_summarylin}).
Overall, the cross section for mono-$h$ in the standard scalar DM Higgs portal is predicted to be very small.

\vspace{2mm}

\begin{figure}[t!]\centering
\includegraphics[width=.7\textwidth]{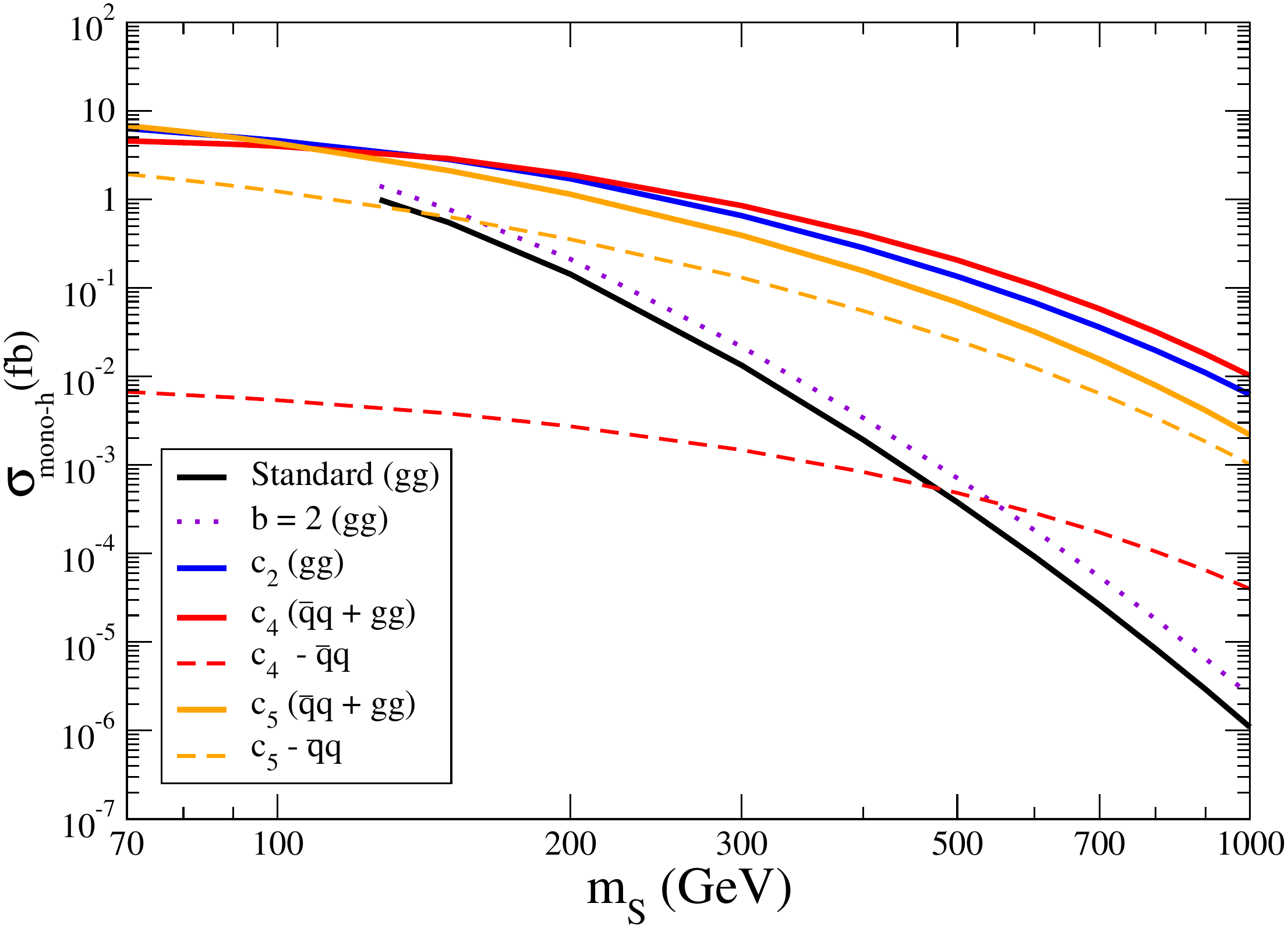}
\caption{Cross section of the process $pp\to h\,SS$ at $\sqrt s=\unit[13]{TeV}$ as a function of $m_S$ for the standard Higgs portal with $\lambda_S = 1$ (solid-black line)
and different non-linear setups. The dotted-purple line corresponds to the case $\lambda_S = 1$, $\bb = 2$, $c_i = 0$. The solid-blue, solid-red and solid-orange lines correspond 
to $\lambda_S = 0$ and $c_2 = 1$, $c_4 = 1$, $c_5 = 1$ respectively. For the latter two cases, the dashed-red and dashed-orange lines show the $\bar{q}q$-initiated 
contribution from $\A_4$ and $\A_5$. The low mass end-point for the solid-black and dotted-purple lines, given by $m_S = 127$ GeV, corresponds to the mass bound for the 
standard Higgs portal scenario for $\lambda_S = 1$ (see Figure \ref{plots_summarylin}).}\label{plot_monoH_xs}
\end{figure}

The presence of  non-linear Higgs-DM interactions may significantly change the previous picture, as the suppression factors for the standard scenario can be 
overcome by the appearance of new production channels -- {\it e.g.}~direct couplings of DM to $Z$-bosons which yield a $\bar{q}q$-initiated mono-$h$ contribution 
(case of $\A_4$ and $\A_5$) --  and/or by the momentum dependence of the $S$-$h$, $S$-$Z$ and $S$-$h$-$Z$ interactions (case of $\A_2$, $\A_4$ and $\A_5$). A sample of the Feynman diagrams 
contributing to mono-$h$ in this case is shown in Figure \ref{diagram_monoHprod}. 
For $\A_2$, mono-$h$ is $g g$-initiated, and the amplitude receives contributions from Feynman diagrams scaling as $\sim c_2$ and $\sim c^2_2$. $\A_4$ and $\A_5$ yield $g g$- 
and $\bar{q}q$-initiated contributions to the mono-$h$ process, both scaling linearly with $c_{4,5}$. In Figure~\ref{plot_monoH_xs} we illustrate the behavior of the 
cross section $\s_{\text{mono}-h}=\s(pp \to h\,SS)$ as a function of the DM mass $m_S$ at a centre of mass ({\it c.o.m.})~energy of $\sqrt{s}=\unit[13]{TeV}$, for
each of the possible non-linear operators $\A_i$ with $c_i = 1$ and $\lambda_S = 0$ compared to the standard Higgs portal with $\lambda_S = 1$ (solid-black line). 
Let us first note that a non-linear value $\bb > 1$ (dotted-purple line) enhances several processes $\sim \lambda_S$ {\it w.r.t.}~the standard Higgs portal scenario
(which modifies the interference between $\sim \lambda_S$ and $\sim \lambda^2_S$ terms) and yields a somewhat larger mono-$h$ cross section. More importantly,
Figure~\ref{plot_monoH_xs} shows that the presence of either of $\A_2$ (solid-blue line), $\A_4$ (solid-red line), $\A_5$ (solid-orange line) 
may lead to a large enhancement in the cross section for DM masses $m_S > 100$ GeV,  potentially reaching enhancements of order $10^4 \times c_i^2$ for $m_S \gg v$ (we recall that $\lambda_S = 1$ for the standard Higgs portal scenario is only allowed 
for $m_S > 127$ GeV, and the same bound applies roughly to the scenario $\bb \neq 1$, as this only has a significant impact on the value of 
$\Omega_S$ for $m_S > m_h$, as shown in Figure \ref{plot_omegab}). 

\vspace{2mm}

\begin{figure}[t!]\centering
\includegraphics[width=.49\textwidth]{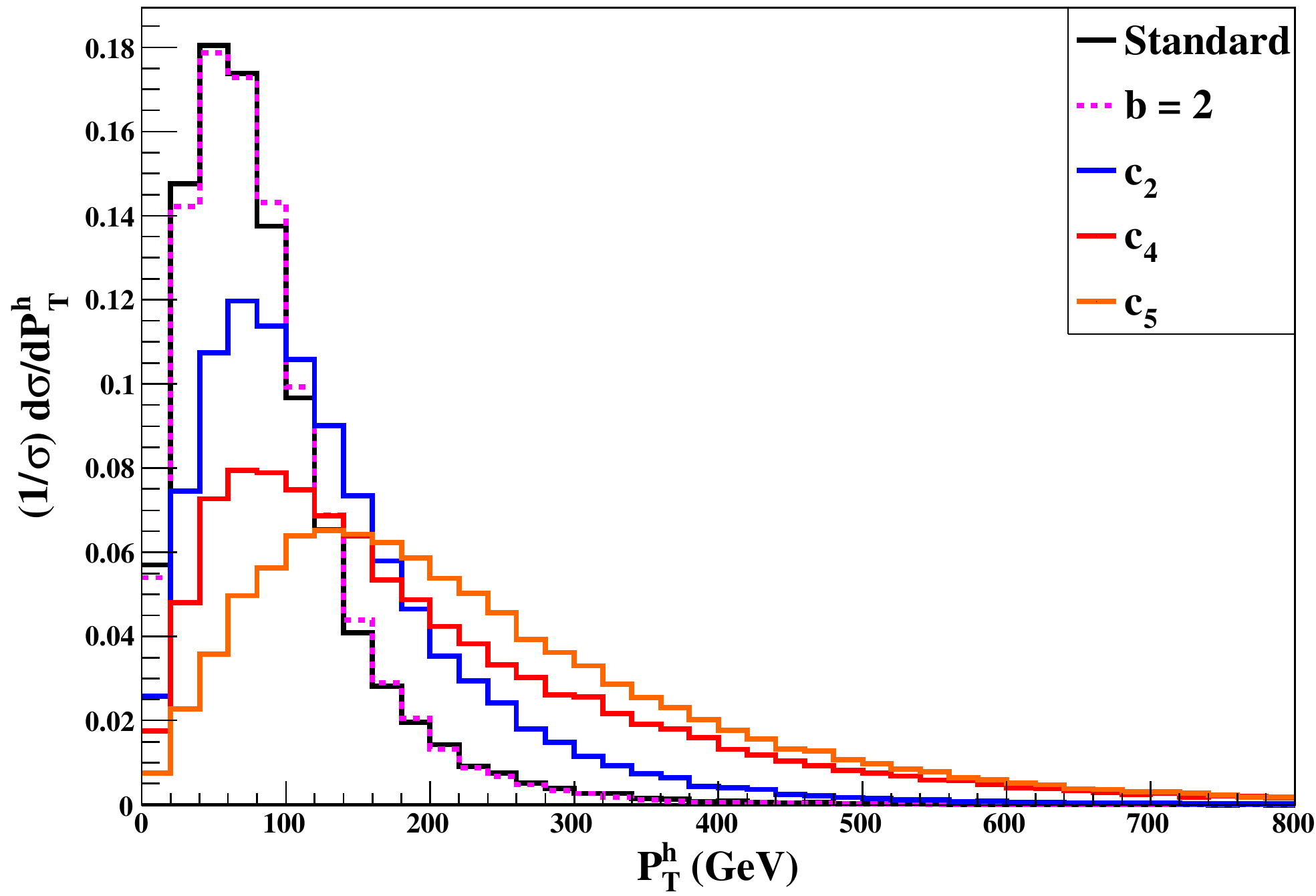}
\includegraphics[width=.49\textwidth]{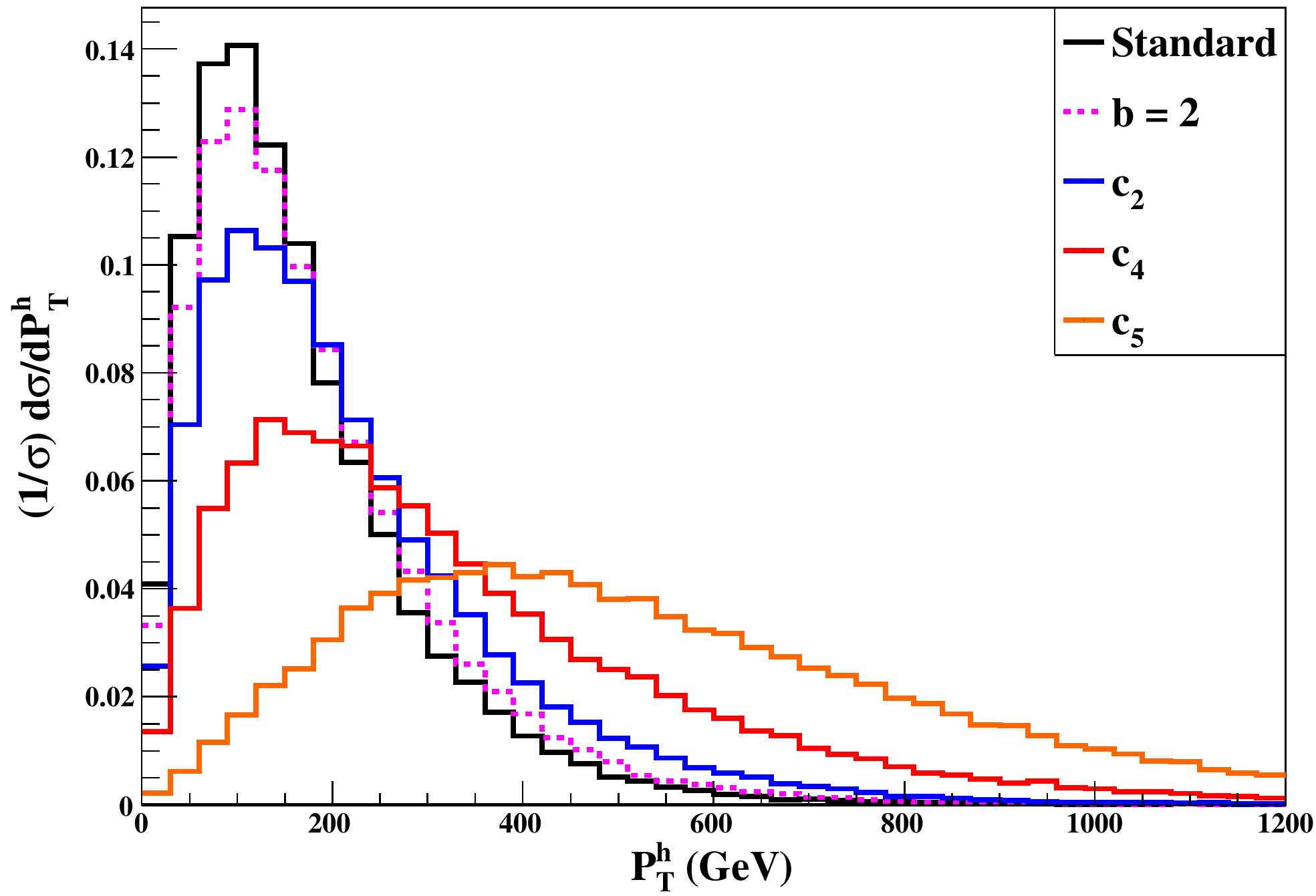}
\caption{Normalised differential $P_T^h$ distribution for the process $pp\to h\,SS$ in the standard Higgs portal with $\lambda_S = 1$ (solid-black line), 
non-linear Higgs portal with $\bb = 2$ (dashed-purple line), $\A_2$ with $c_2 = 1$ (solid-blue line), $\A_4$ (solid-red line) and $\A_5$ (solid-orange line), 
for $m_S = 100$ GeV (Left) and $m_S = 500$ GeV (Right).}\label{Diff_monoH}
\end{figure}

Besides the potentially large increase in the mono-$h$ cross section, in the presence of $\A_{2,4,5}$ the differential distribution 
of the Higgs boson transverse momentum $P^h_T$ is shifted towards larger values, as shown in Figure \ref{Diff_monoH} for $m_S = 100$ GeV (Left)
and $m_S = 500$ GeV (Right). This much harder mono-$h$ $P^h_T$ spectrum, particularly for the case of $\A_5$, is a landmark
signature of non-linear Higgs portals, which also allows for a much better signal extraction from the SM background. 

\vspace{2mm}

Finally, let us stress that given the 13 TeV results from Figure \ref{plot_monoH_xs} the 8 TeV mono-Higgs searches at the LHC do not 
put any meaningful constraint on the parameter space under discussion here, since if we assume a SM value for Br($h\to \gamma\gamma$) $\simeq 2\cdot 10^{-3}$ the ATLAS 
95\% C.L. limit \cite{Aad:2015yga} on the fiducial mono-$h$ cross section is $\s_{\text{mono-$h$}}\leq \unit[0.35]{pb}$, two orders of magnitude larger than the (13 TeV) 
cross sections showed in Figure \ref{plot_monoH_xs}.

\subsubsection{Mono-Z and mono-W searches}

As a last category of DM observables, we discuss the searches for mono-$W^\pm$~\cite{Bai:2012xg} and 
mono-$Z$~\cite{Bell:2012rg,Carpenter:2012rg,Alves:2015dya,Neubert:2015fka} signatures at the LHC in the context of Higgs portal scenarios. 
We first focus on the process $pp \to Z S S$, which receives non-linear contributions from all the effective operators $\A_i$ in Eq.~\eqref{scalar.op}. 
Both for the standard Higgs portal scenario and in the presence of $\A_1$, $\A_2$, $\A_3$, $\A_4$ these contributions are both 
$g g$- and $\bar{q}q$-initiated, while $\A_5$ only gives rise to $g g$-initiated contributions to mono-$Z$.
We also note that $\A_1$ and $\A_3$ give exactly the same contribution to the mono-$Z$ process if $c_1 = 2\,c_3$ - see Appendix~\ref{Feynman_rules}, 
and furthermore $c_3 \lesssim 0.1$ is 
required from EW precision data (recall the discussion at the end of Section \ref{Sect:Nonlinear}), 
so in the following we do not explicitly discuss the impact of $\A_3$ on mono-$Z$ searches. 

\vspace{2mm}

\begin{figure}[t!]\centering
\includegraphics[width=.49\textwidth]{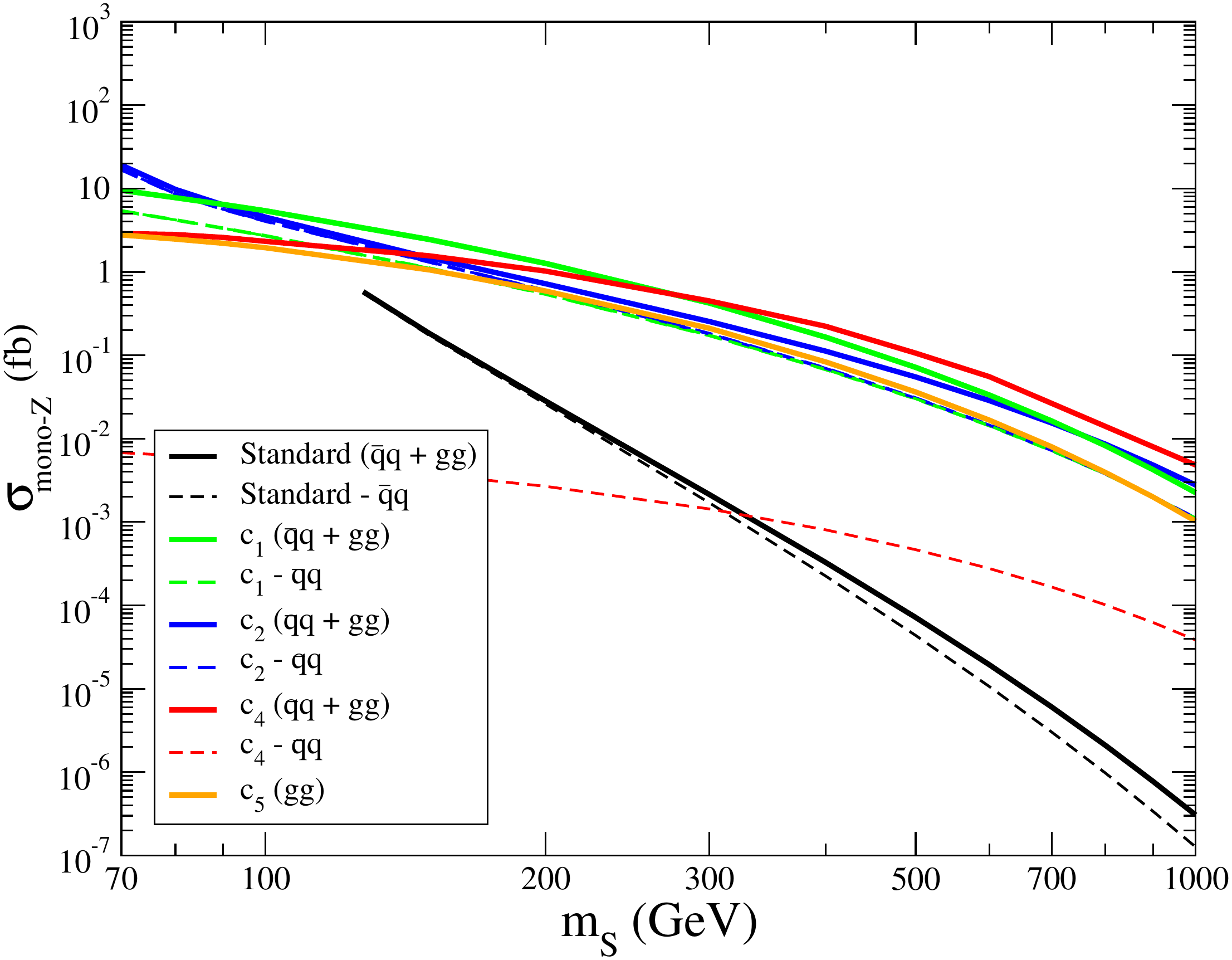}
\includegraphics[width=.49\textwidth]{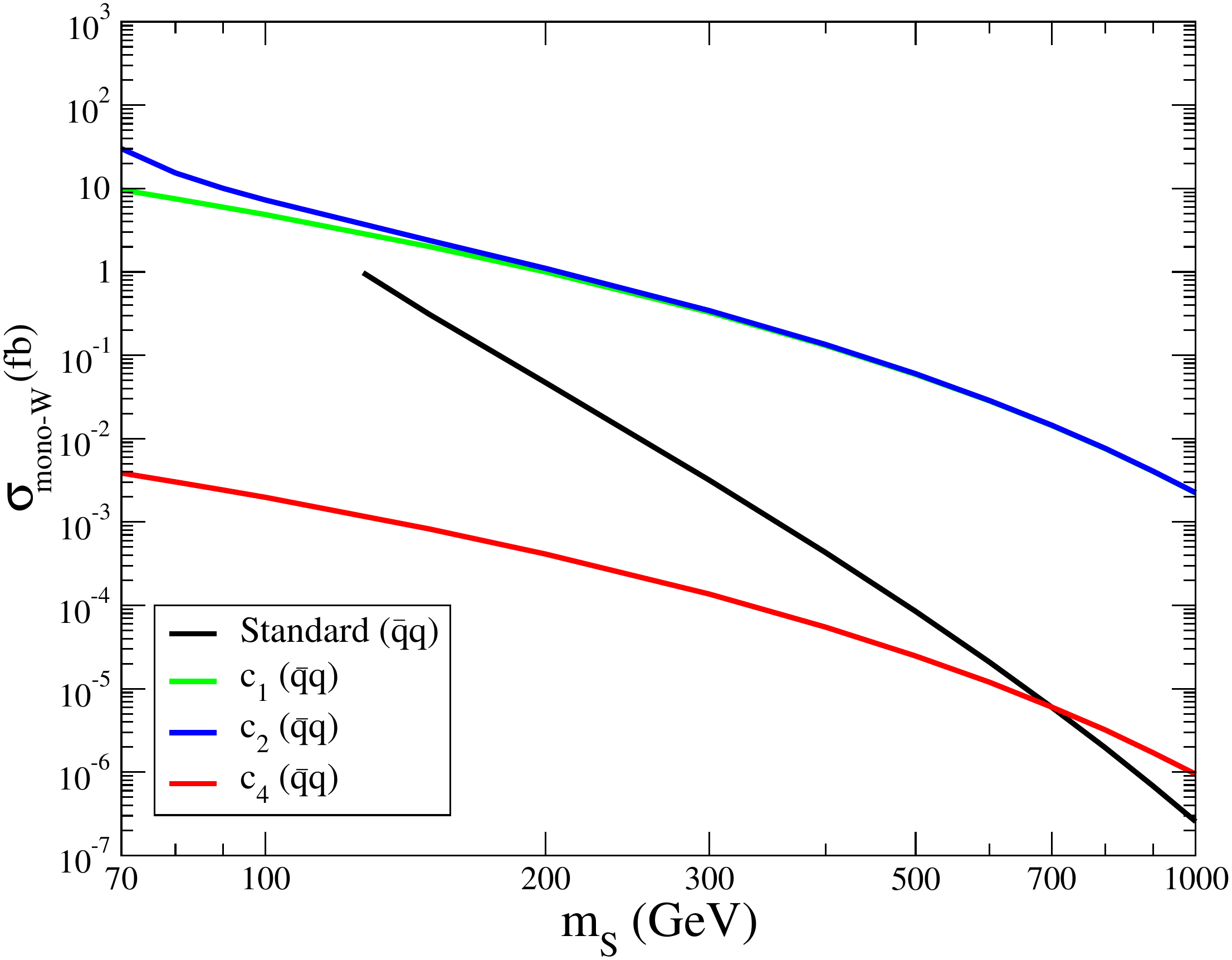}
\caption{Cross section of the process $pp\to Z\,SS$ (Left) and $pp\to W^{\pm}\,SS$ (Right) at $\sqrt s=\unit[13]{TeV}$ as a function of $m_S$ for 
the standard Higgs portal with $\lambda_S = 1$ (solid-black line) and different non-linear setups. 
The solid-green, solid-blue, solid-red and solid-orange lines correspond to $\lambda_S = 0$ and $c_1 = 1$, $c_2 = 1$, $c_4 = 1$, $c_5 = 1$ respectively.
In the Left Figure, the dashed-black, dashed-green, dashed-blue and dashed-red lines respectively show the $\bar{q}q$-initiated contribution to the process $pp\to Z\,SS$ for 
the standard, $\A_1$, $\A_2$ and $\A_4$ scenarios.}\label{plot_monoWZ_xs}
\end{figure}

\begin{figure}[t!]\centering
\includegraphics[width=.49\textwidth]{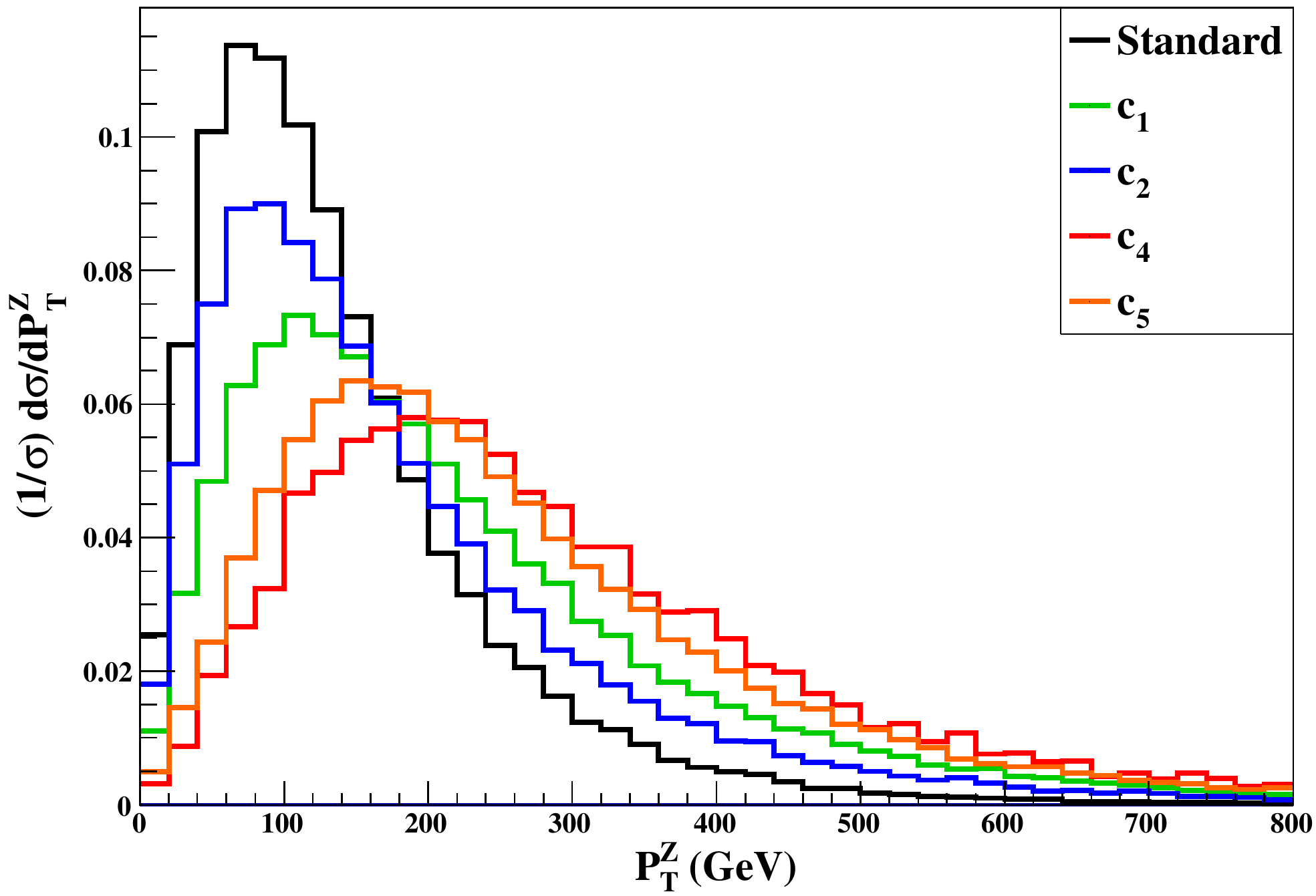}
\includegraphics[width=.49\textwidth]{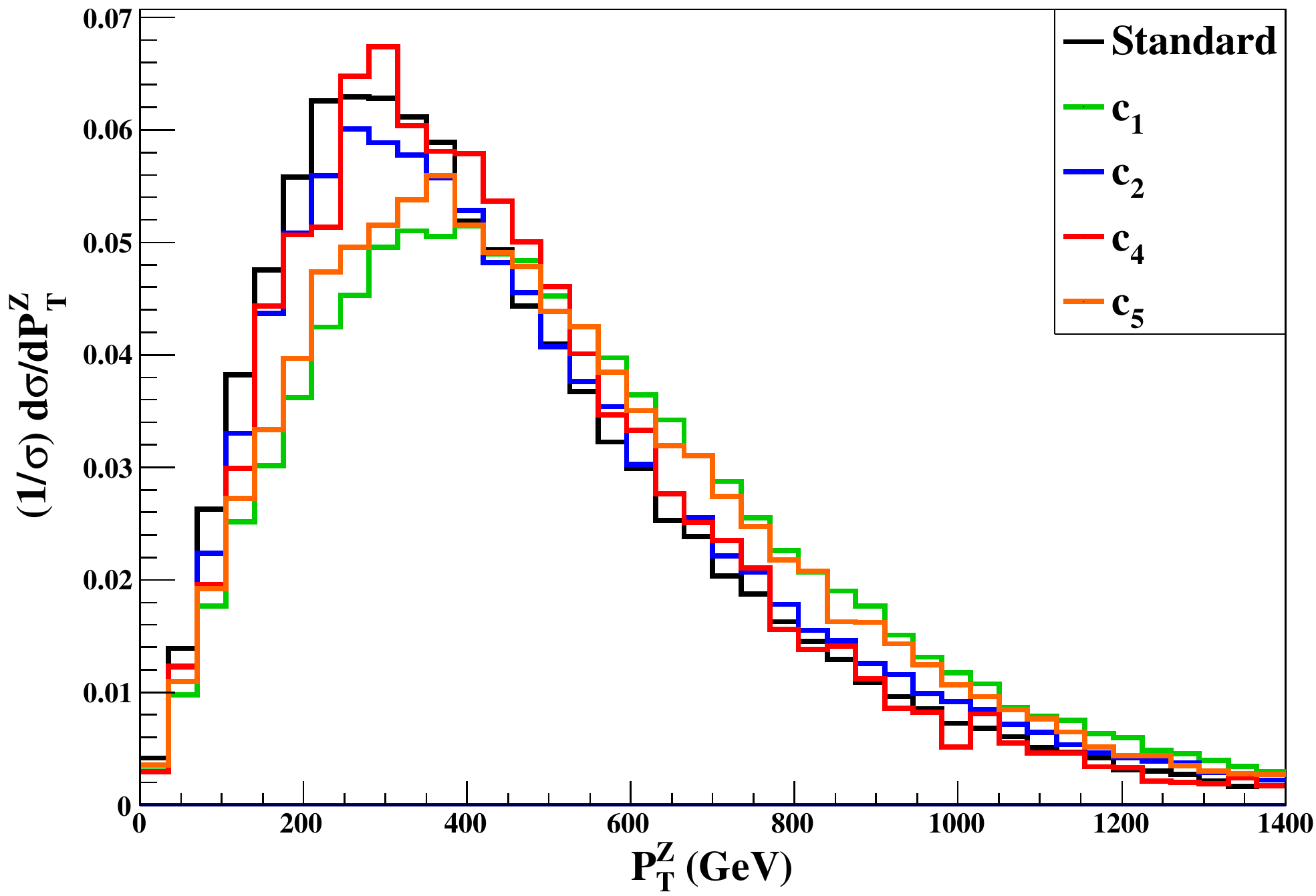}
\caption{Normalised differential $P_T^Z$ distribution for the process $pp\to Z\,SS$ in the standard Higgs portal with $\lambda_S = 1$ (black line), 
and for non-linear Higgs portal operators $\A_1$ (green line), $\A_2$ 
(blue line), $\A_4$ (red line) and $\A_5$ (orange line), for $m_S = 100$ GeV (Left) and $m_S = 500$ GeV (Right).}\label{Diff_monoZ}
\end{figure}

In Figure~\ref{plot_monoWZ_xs} (Left) we show the LHC cross sections $\s(pp\to ZSS)$ as a function of $m_S$ for a {\it c.o.m.}~energy $\sqrt{s}=\unit[13]{TeV}$. 
The solid-black line corresponds to the standard Higgs portal scenario with $\lambda_S=1$ (with $\s^{\mathrm{standard}}_{\mathrm{mono}-Z} \sim \lambda^2_S$), which
decreases quite fast for increasing $m_S$. As in the mono-$h$ case (see Section \ref{Section_monoH}), the solid-green, solid-blue, solid-red and solid-orange curves
respectively correspond to non-linear Higgs portal scenarios with $\lambda_S=0$ and $\A_1$, $\A_2$, $\A_4$ or $\A_5$ being present with $c_i = 1$. In all the non-linear setups, $\s_{\text{mono-}Z}^i\sim c_i^2$, the only exception being $\A_4$, which contributes with diagrams scaling both as $c_4$ and as $c_4^2$. 
As can be seen from Figure~\ref{plot_monoWZ_xs}, these non-linear contributions yield a significantly larger mono-$Z$ cross section as compared to the standard Higgs portal for $m_S \simeq 100$ GeV, leading to very large enhancements for $m_S \gg v$. As with the mono-$h$ signature, 
the non-linear operators $\A_{1,2,4,5}$ also affect the differential distribution 
of the $Z$-boson transverse momentum $P^Z_T$, yielding a harder mono-$Z$ $P^Z_T$ spectrum, as can be seen from Figure \ref{Diff_monoZ}. 
This effect is more important for DM masses in the range $100 - 300$ GeV, while for $m_S \gg v$ the standard and non-linear $P^Z_T$ spectra become very similar. 
Mono-$Z$ signatures therefore constitute a promising probe of non-linear Higgs portals at the 13 TeV run of the LHC for intermediate DM masses ($m_h/2 < m_S \ll 1$ TeV) and sizable 
values of the coefficients $c_i \lesssim 1$. On the other hand, current 8 TeV mono-$Z$ searches at the LHC
are only able to constrain values $c_i \gg 1$: the ATLAS analysis \cite{Aad:2014vka}, using 20.3 fb$^{-1}$
of LHC 8 TeV data, yields 95\% C.L. limits on the mono-$Z$ ($Z \to \ell^+ \ell^-$) fiducial cross section
$\s^{\ell\ell}_{\text{mono-$Z$}}\leq \unit[2.7]{fb}$, $\unit[0.57]{fb}$, $\unit[0.27]{fb}$, $\unit[0.26]{fb}$ after a corresponding
selection $\slashed{E}_T > 150$ GeV, $250$ GeV, $350$ GeV, $450$ GeV. Such limits lie well above the (13 TeV) curves in Figure ~\ref{plot_monoWZ_xs} (Left), and moreover 
for fairly light DM ($m_S \lesssim 100$ GeV) the selection criteria from the ATLAS search \cite{Aad:2014vka} will discard most of the DM signal, as shown in Figure \ref{Diff_monoZ}. 

\vspace{2mm}

\begin{figure}[t!]\centering
\includegraphics[width=.49\textwidth]{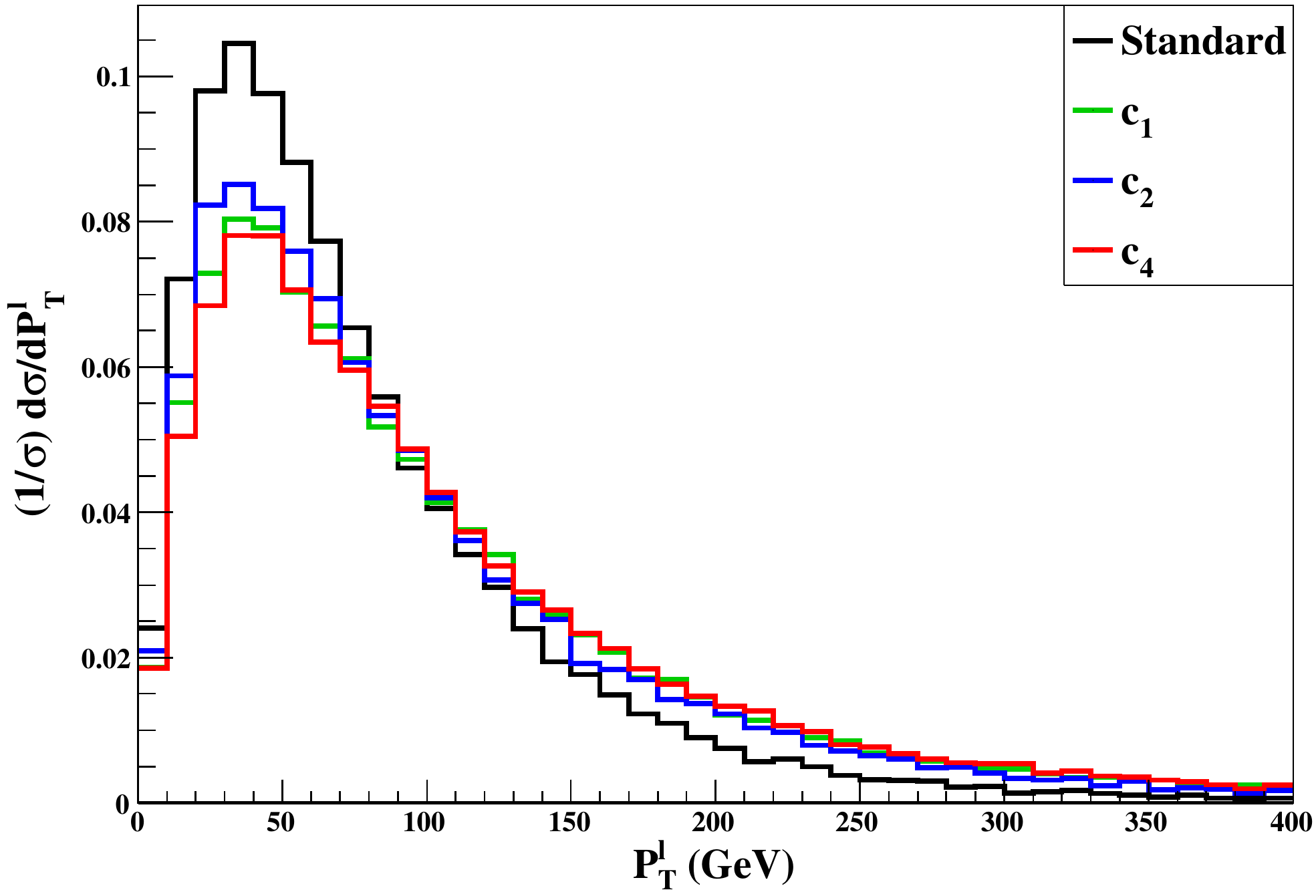}
\includegraphics[width=.49\textwidth]{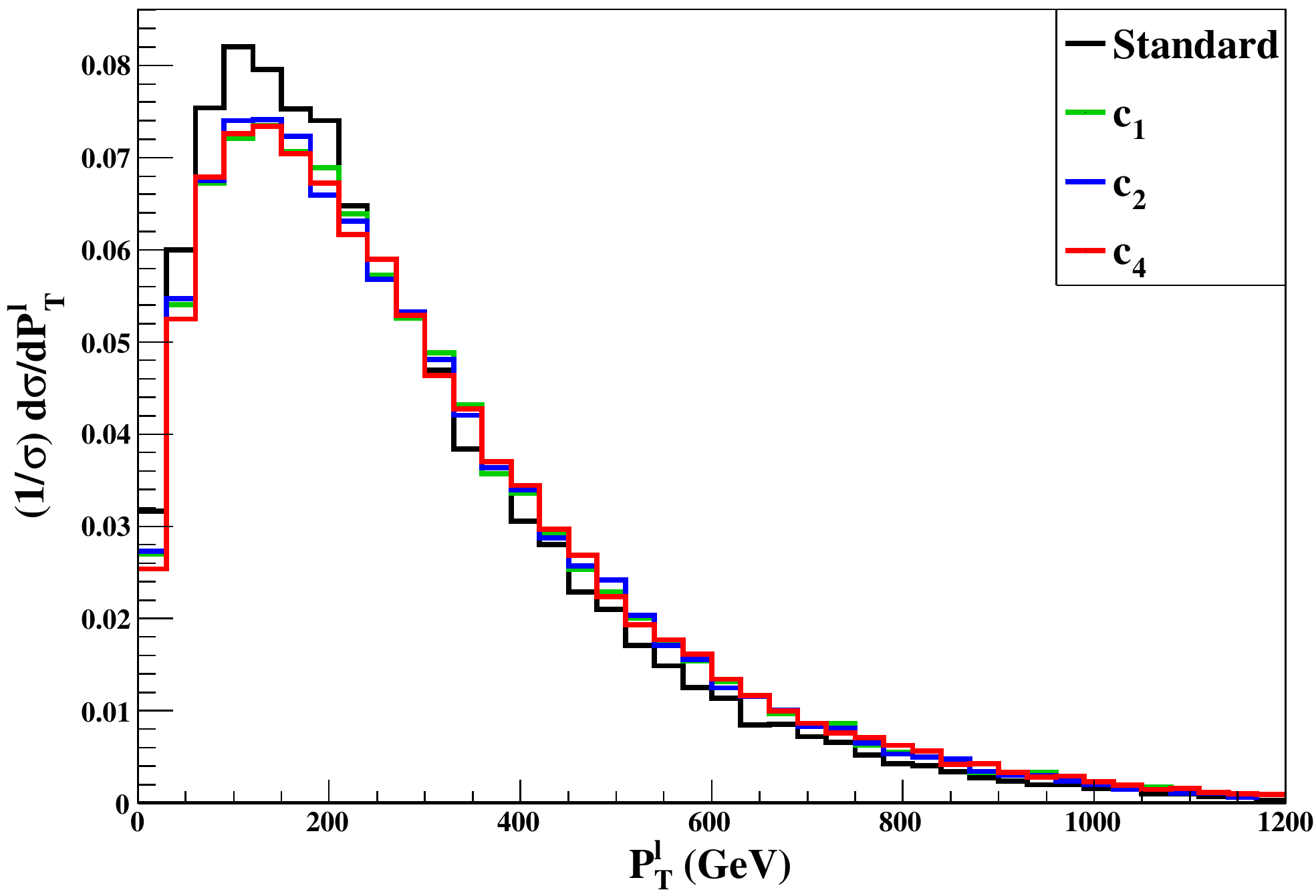}
\caption{Normalised differential $P_T^{\ell}$ distribution for the process $pp\to W^{\pm}\,SS$ ($W^{\pm} \to \ell^{\pm}\,\nu_{\ell}$) in the 
standard Higgs portal (black line), and for non-linear Higgs portal operators $\A_1$ (green line), $\A_2$ 
(blue line) and $\A_4$ (red line), for $m_S = 100$ GeV (Left) and $m_S = 500$ GeV (Right).}\label{Diff_monoW}
\end{figure}

Turning now to mono-$W^\pm$ signatures, these are affected by the non-linear operators $\A_1$, $\A_2$ and $\A_4$. Both for these operators and for the
standard Higgs portal, the contributions to mono-$W^\pm$ are all $\bar{q}q$-initiated, which as we will see makes an important difference {\it w.r.t.} the case of 
mono-$Z$ signatures. In Figure~\ref{plot_monoWZ_xs} (Right) we show the cross section $\s(pp\to W^{\pm} SS)$ as a function of $m_S$ for the standard and non-linear 
Higgs portal scenarios (using the same criteria and colour convention as for the mono-$Z$ analysis). In the presence of $\A_1$ and/or $\A_2$ a significant enhancement in the 
cross section can occur for large values of $m_S$, similar to the case of mono-$Z$ and mono-$h$ signatures. However, for the operator $\A_4$ mono-$W^{\pm}$ signatures are very 
suppressed, as the dominant $g g$-initiated contribution (compare the solid- and dashed-red lines in Figure~\ref{plot_monoWZ_xs} (Left) for mono-$Z$) 
is absent in this case. We find that, contrary to the situation encountered in the mono-$h$ and mono-$Z$ analyses above, for mono-$W^\pm$ signatures with 
$W^{\pm} \to \ell^{\pm}\,\nu_{\ell}$ the $P^{\ell}_T$ of the final state lepton has a very similar distribution for the standard and non-linear Higgs portal scenarios, both 
for low and high DM masses, as seen in Figure \ref{Diff_monoW}.

\vspace{2mm}

\begin{figure}[t!]
\begin{subfigure}{.495\textwidth}\centering
\hspace*{-0.5cm}
\includegraphics[height=5.9cm]{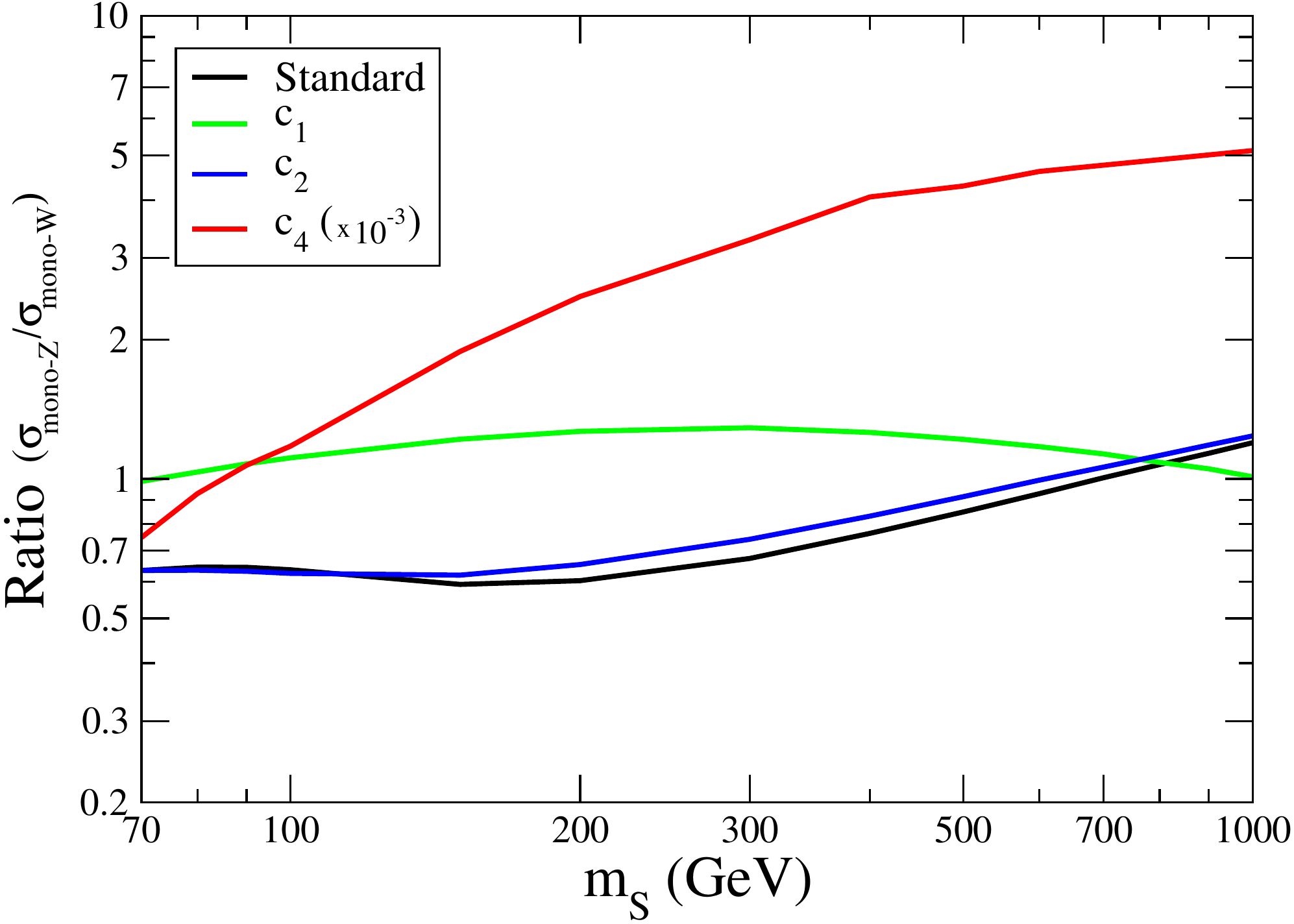}
\end{subfigure}
\begin{subfigure}{.495\textwidth}\centering
\includegraphics[trim = 0cm 0cm 0cm 0cm, clip=true, totalheight=5.9cm]{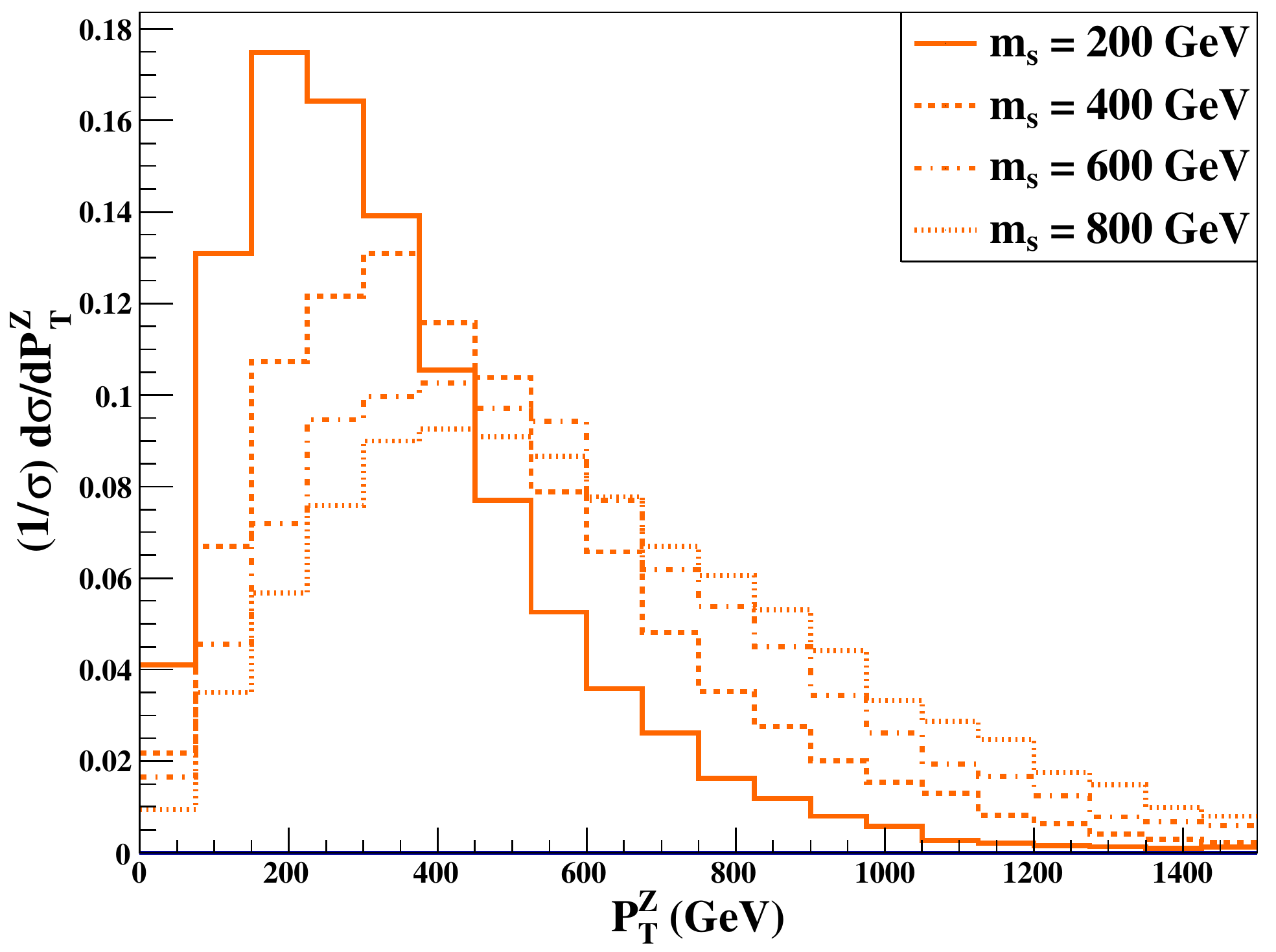}
\end{subfigure}
\caption{Left: Cross section ratio $R_{WZ} \equiv \s(pp\to Z SS)/\s(pp\to W^{\pm} SS)$ at $\sqrt s=\unit[13]{TeV}$ as a function of $m_S$ 
 in the standard Higgs portal scenario (black line) and for the non-linear operators $\A_1$ (green-line), $\A_2$ (blue-line) and $\A_4$ (red-line), the latter ratio having been 
 multiplied by $10^{-3}$ to be shown in the Figure. Right: Normalised differential $P_T^Z$ distributions for the process $pp\to Z\,SS$ 
for $\A_5$ and DM masses $m_S = 200$ GeV (solid), $400$ GeV (dashed), $600$ GeV (dash-dotted) and $800$ GeV (dotted).}\label{plot_monoZ_vs_monoW} 
\end{figure}

Finally, we discuss the possibility of using the ratio $R_{WZ} \equiv \s(pp\to Z SS)/\s(pp\to W^{\pm} SS)$ as a probe of non-linear Higgs portal scenarios, 
as shown in Figure~\ref{plot_monoZ_vs_monoW} (Left) as a function of $m_S$. Remarkably, the impact of each non-linear operator on this ratio is determined only by its gauge and Lorentz structure, independently of the value of the coefficient\footnote{The line for $\A_4$ is an exception, due to the fact that the amplitude for mono-$Z$ receives contributions scaling both as $c_4$ and as $c_4^2$, so that the coefficient does not fact out in $R_{WZ}$. However, this does not impair the interpretation of the plot in Fig.~\ref{plot_monoZ_vs_monoW}.} $c_i$. Analogously, the dependence on $\lambda_s$ factors out in the standard case. While the effect of the operator $\A_2$ on this observable cannot be effectively disentangled from that of a standard Higgs portal 
 (as can be seen by comparing the black and blue curves in Figure~\ref{plot_monoZ_vs_monoW} (Left)), 
 the ratio $R_{WZ}$ is a very powerful non-linear discriminator for the cases of $\A_1$ and $\A_4$ (also trivially for $\A_5$, for which the mono-$W^{\pm}$ process is absent and $R_{WZ}\equiv\infty$), corresponding 
respectively to the green and red curves in Figure~\ref{plot_monoZ_vs_monoW} (Left). 
Moreover, recalling that the operator $\A_3$ enters the mono-$Z$ process with the corresponding coefficient in the 
combination $(c_1+2c_3)$ (see Appendix \ref{Feynman_rules}), while it does not enter the mono-$W^{\pm}$ process, the green curve in 
Figure~\ref{plot_monoZ_vs_monoW} (Left) will get rescaled by $(c_1+2c_3)^2/c^2_1$ in the presence of $\A_3$. Thus, for $sign(c_1) = sign(c_3)$, 
the green curve actually represents  a lower bound on the contribution of $\A_1$ and $\A_3$ to the ratio $R_{WZ}$. 

Importantly, it is in principle possible to infer the DM mass from the mono-$Z$/mono-$W^{\pm}$ processes through the differential information on the $P_T^{V}$ ($V = W^{\pm},Z$) as shown explicitly in Figure~\ref{plot_monoZ_vs_monoW} (right) for the case of $\A_5$ (alternatively, the $\slashed{E}_T$ distribution may be used).
Taking this into account, the hypothetical observation of mono-$Z$ and mono-$W$ signals would allow to extract at the same time a measurement of  $R_{WZ}$ and of $m_S$, \textit{i.e.} to identify a unique point (surrounded by a finite error region) in the parameter space of figure~\ref{plot_monoZ_vs_monoW} (Left). Naively, the further this point lies away from the black line, the more disfavored the standard portal scenario will be. Employing this technique in a more thorough analysis, which would keep all the relevant uncertainties into account, it would be possible to quantify a confidence level for the exclusion of the standard portal. Therefore, the ratio $R_{WZ}$ can be an efficient probe of the nature of the DM portal to the SM.  Notice that the non-linear scenario cannot be ruled out by this kind of study, since any point in the $(m_S,R_{WZ})$ space corresponds to a whole set of combinations of the coefficients $c_{1-5}$.

\subsection{A comment on indirect detection of Dark Matter}
DM annihilation into charged particles (or states further decaying into charged particles), whether $W\pm$ or charged fermions, 
would result in significant fluxes of gamma-rays, which can be constrained by astrophysical observations, {\it e.g.}~from the Fermi-LAT Space Telescope.
Rather than performing a detailed study of the indirect detection signatures of non-linear Higgs portal DM scenarios (which we defer for the future), 
we just discuss briefly the impact of such indirect limits on their parameter space, focusing on DM annihilation into $W^{+}W^{-}$ 
and $b\bar{b}$, which receive contributions from $\A_{1,2,4}$ and $\A_{2,4}$ respectively (see Appendix \ref{Relic_Abundance}).
We consider the limits on such DM annihilation channels from measurements of the gamma-ray flux from the Milky Way galactic center \cite{Vitale:2009hr}, 
which have been shown to be competitive \cite{Hooper:2012sr} with those derived from other astrophysical sources, such as dwarf galaxies. 
Using the limits from \cite{Hooper:2012sr} on the DM annihilation cross-section $(\s \vel)_\text{ann}$ into $W^{+}W^{-}$ 
and $b\bar{b}$, given respectively by Eqs.~(\ref{SSWW}) and (\ref{SSff}), we can potentially derive constraints on $\lambda_S$ and/or $c_i$ as a function of the DM 
mass $m_S$. After the appropriate rescaling of the indirect DM signal by $(\Omega_{S}/\Omega_{\mathrm{DM}})^2$, we find that the current limits from \cite{Hooper:2012sr} do not 
provide a meaningful constraint on the parameter space under consideration.


\section{Connection with the linear EFT expansion}
\label{EFT_DM_Linear}
In this section the connection between the non-linear scenario analysed in the previous sections and the linear context is 
discussed. Eq.~(\ref{SMHportal}) accounts for the only possible renormalisable coupling between the elementary SM Higgs particle and a 
singlet scalar DM particle (assuming $Z_2$ symmetry). Nevertheless, scenarios for BSM electroweak physics can - and often do - correspond to linear 
realisations of the EWSB mechanism, typical of perturbative completions. A model-independent  parametrisation of the new physics for the SM degrees of freedom 
is then given by higher-dimension operators of mass dimension $\ge4$, suppressed by inverse powers of the BSM physics scale $\Lambda \gg v$ : a linear operator expansion, 
in which the $h$ participates via $\Phi$ insertions and thus through  a $(v+h)$ functional dependence. The question then arises of the extent up to which the signals determined 
above for the non-linear DM portal could be mimicked by effective couplings of the linear expansion, that is by Eq.~(\ref{SMHportal}) plus a tower of operators of mass 
dimension 6, 8 etc.

\begin{table}[ht!]\centering
\renewcommand{\arraystretch}{1.5}
\begin{tabular}{*2{>{$}c<{$}@{ $\longrightarrow$ }>{$}c<{$}@{ $\equiv$ }>{$}l<{$}}}
\multicolumn{3}{c}{$d=6$}& \multicolumn{3}{c}{$d=8$}\\\midrule
\bb & \cO_\bb & (\Phi^\dag \Phi)^2 S^2 &
\A_3& \cO_3 & (\Phi^\dag \stackrel{\leftrightarrow}{D_\mu} \Phi)(\Phi^\dag\stackrel{\leftrightarrow}{D^\mu} \Phi) S^2\\
\A_1& \cO_1 & D_\mu \Phi^\dag D^\mu \Phi\, S^2&
\A_5& \cO_5 & (\Phi^\dag \stackrel{\leftrightarrow}{D_\mu} \Phi) D^\mu \left(\Phi^\dag \Phi\right) S^2\\
\A_2& \cO_2 & \square\left(\Phi^\dag  \Phi\right) S^2&\multicolumn{3}{c}{}\\
\A_4& \cO_4 & (\Phi^\dag \stackrel{\leftrightarrow}{D_\mu} \Phi)D^\mu S^2\hspace*{5mm}
\end{tabular}
\caption{Linear siblings of the non-linear operators $\A_i$ and of the deviations of the standard Higgs portal coupling.}
\label{linearsiblings}
\end{table}
First of all, the couplings of the non-linear Higgs portal, that is, the deviations from the standard portal given by $\bb\ne1$  in Eq.~(\ref{NLHportal}) as well as the operators $ \A_1-\A_5$,  appear among the dominant couplings of that expansion, while their linear counterparts are not found at the renormalisable level but only at higher orders in the expansion. Indeed, the siblings (lowest dimension operators in the linear expansion which contain at least the same physical couplings) of $\A_1$, $\A_2$, $\A_4$ and  the linear operator inducing $\bb\ne1$ are  linear operators of mass dimension $d=6$, while the couplings $\A_3$ and $\A_5$ would first appear as $d=8$ linear operators. The explicit definition of the linear siblings can be found in Table~\ref{linearsiblings}, providing a one-to-one mapping between the linear and the non-linear operators. 

The complete $d=6$ bosonic linear portal describing the interaction with at most two $S$ fields includes, in addition to  $\cO_1$, $\cO_2$, $\cO_4$ and $\cO_b$ above,  9 four-derivative couplings~\footnote{Other bosonic operators are redundant in that they are related via equations of motion or a total derivative; for instance the operator $\de_\mu S \de^\mu S\Phi^\dag \Phi$  can be reabsorbed by field redefinitions.}:
\begin{equation}
\begin{aligned}
&g^2 S^2 W_{\mu\nu}W^{\mu\nu}\qquad\qquad\qquad
&&g^2 S^2 W_{\mu\nu}\tilde{W}^{\mu\nu}\\
&g^{\prime2} S^2 B_{\mu\nu}B^{\mu\nu}
&&g^{\prime2} S^2 B_{\mu\nu}\tilde{B}^{\mu\nu}\\
&gg' S^2 B_{\mu\nu}W^{\mu\nu}
&&gg' S^2 B_{\mu\nu}\tilde{W}^{\mu\nu}\\
&g_s^2 S^2 G_{\mu\nu}G^{\mu\nu}
&&g_s^2 S^2 G_{\mu\nu}\tilde{G}^{\mu\nu}\\
&\square S\square S
\label{fourderiv}
\end{aligned} 
\end{equation}
Being four-derivative couplings, these operators would correspond to sub-dominant operators in the non-linear expansion considered here, which includes at most two-derivative operators; they will thus be disregarded in what follows.

As in the case of non-linear expansion, in order to define a complete basis, fermionic structures should be also considered in addition to those in Eq.~(\ref{Fermionic2}):
\begin{equation}
\bar{Q}_{L_i}\Phi\, d_{R_j} S^2\,,\qquad
\bar{Q}_{L_i}\tilde\Phi\, u_{R_j} S^2\,,\qquad 
\bar{L}_{L_i}\Phi\, e_{R_j} S^2\,.
\label{Fermionic1Linear}
\end{equation}
Again, two flavour blind combinations of the two types of chiral fermion structures (Eq.~(\ref{Fermionic2}) and (\ref{Fermionic1Linear})) may be related to the bosonic operators $\cO_2$ and $\cO_4$, respectively. In order to avoid redundancies either the two combination or the two latter bosonic operators should then be disregarded~\cite{UsFuture}.

The sector of the linear effective Lagrangian containing the siblings of interest for the comparison is then given by
\begin{equation}
\label{linearportal}
\LL^\text{linear portal}_S\supset \sum_{i=\bb,1,2,4} \dfrac{c^L_i}{\Lambda_{DM}^2} \cO_i+\sum_{i=3,5} \dfrac{c^L_i}{\Lambda_{DM}^4} \cO_i\,,
\end{equation}
where $c^L_i$ denote the operator coefficients.

The rationale of the operator expansions calls for their dimensionless parameters to be naturally $\cO (1)$, in which case the answer to the question formulated above is obvious: while $\A_1-\A_5$ may be expected to contribute with similar strength to the couplings in Eq.~(\ref{NLHportal}), the $d \ge 6$ operators 
of the linear expansion should be suppressed by powers of $v^2/\Lambda_{DM}^2 \ll 1$: in other words, the dominant, leading order effects of the linear expansion are expected to reduce exclusively to those of the standard portal in Eq.~(\ref{SMHportal}), in contrast to the plethora of phenomenological consequences of the leading-order non-linear portal. 

It could be argued, though, that fine-tunings occur in nature: in a particular model the amplitude of a given leading operator of the linear expansion could be suppressed, or alternatively that of a higher-dimension operator enhanced. In such an hypothetical situation, is there a way to disentangle the origin of a putative signal of the non-linear basis with respect to that from a sibling linear operator? The answer is positive even if the procedure is involved: a further tool is provided by the comparison -- for a given type of coupling -- between a vertex with no $h$ leg versus one or more additional $h$ legs, because they are correlated in the linear case and not so in the non-linear one. For instance the Feynman Rules in Appendix~\ref{Feynman_rules}, and in particular FR.2 {\it vs.} FR.6, illustrate that the couplings  $S-S-Z$ and $S-S-Z-h$ are correlated. This is not the case in the non-linear scenario, where these couplings are independent of one another. 
An analogous effect, due to the different orderings of the operators in the two expansions, is visible in FR.4 {\it vs.} FR.5: whilst the vertices $S-S-W-W$ and $S-S-Z-Z$ are proportional to each other in the linear description, they are no longer so in the non-linear case.
In practice, such an analysis would be challenging from the experimental point of view, as the identification of these specific couplings is not straightforward with the observables considered here. 

Note finally that while some apparent decorrelation may still happen  in the linear expansion via a fine-tuned combination of couplings of different orders, with enough data on Higgs physics a global analysis should provide enough resolution on the nature of EWSB involved. Furthermore, that nature would also be expected to show up in other BSM couplings not involving the DM particle.

\vspace{0.2cm}

On a different realm, notice that the comparison between the non-linear portal and the $d\ge 6$ in Eq~(\ref{linearportal})  implies a trivial relation between the Lagrangian coefficients of the two expansions, when comparing the intensity of the interactions:
\begin{equation}
c^L_i\dfrac{v^2}{\Lambda_{DM}^2}=c_i\qquad \text{for}\quad i=1,2,4\,,\qquad\qquad
c^L_i\dfrac{v^4}{\Lambda_{DM}^4}=c_i\qquad \text{for}\quad i=3,5\,.
\label{EquivalenceCoefficients}
\end{equation}
It is then straightforward to apply to the linear analysis the results in the plots presented in the previous sections for the non-linear scenario. A caveat should be kept in mind, though, given the limits of validity of the linear expansion: because $v/\Lambda_{DM}\ll 1$, only those examples explored in which the constraint imposed on the analysis translates into a non-linear coefficient $c_i^L<4\pi$, and  within the region  $\Lambda_{DM}>m_S$,  should be retained for consistency of the perturbative expansion,  as far as no extra exotic light resonance is detected.

Furthermore, note that in the decoupling limit of the two expansions, $\Lambda\to\infty$ (corresponding to $c_i\to0$ in the non-linear case), the effects of the operators $\A_i(h)$ (and of the $\bb\ne1$ deviations) as well as of their linear siblings vanish. Equivalently, the profiles in the figures in the previous sections approach the standard linear DM portal as the values of the coefficients $c_i$ (and of the $\bb$ deviation) get smaller. This can be explicitly seen in Fig.~\ref{plots_summaryc2}, where the excluded parameter space increases with the coefficient $c_1$ getting smaller in absolute value.


\clearpage
\section{Discussion and conclusions}
\label{Sect:Concl}

In this paper we have studied a new, more general scenario of scalar Higgs portals, with electroweak symmetry breaking non-linearly realised. 
Within this pattern of symmetry breaking, the physical Higgs particle does not behave as an exact  $SU(2)_L$ doublet and in general its participation in couplings as powers of $v+h$ -characteristic of the SM and also of BSM linear realisations of the Higgs mechanism- breaks down.  We have first noticed how this fact automatically transforms the standard scalar Dark Matter Higgs portal and impacts strongly  on the relic abundance.  We have then comprehensively described the non-linear Higgs portal to Dark Matter: the dominant effective couplings -- those not explicitly suppressed by any beyond the SM scale -- describing the interactions of a scalar singlet Dark Matter particle with the Higgs field when electroweak symmetry is non-linearly realised.  A plethora of new couplings appear involving the SM bosonic sector.  The new interactions are characterised by
\begin{itemize}
  \item {\bf Direct couplings to gauge bosons:  } Dark Matter couples to all {\it Higgs degrees of freedom}, namely the Higgs and the longitudinal $W^\pm$ and $Z$, see Eqs.~\ref{Udef}, \ref{oldchiral} and \ref{scalar.op}. 
  \item {\bf De-correlation of single and double Higgs couplings: } The strength of Dark Matter couplings to one- and two-Higgs fields are are de-correlated in non-linear EWSB, see Eq.~\ref{L0S}.
  \item {\bf Novel kinematic features: } Non-trivial momentum dependence of Dark Matter interactions due to new derivative couplings provides handles to disentangle linear vs non-linear Higgs portals at colliders. These features can be extracted from the Lagrangian Eq.~\ref{scalar.op}, and the Feynman rules derived in Appendix~\ref{Feynman_rules}.
\end{itemize}

We have exploited the features of non-linear Higgs portals using information from CMB measurements, Dark Matter direct detection experiments and LHC searches of visible objects recoiling against missing energy. The effect of non-linear interactions on these observables is summarised in Table~\ref{tab:contributions}.

As a general feature, in presence of non-linearity the space of parameters for Higgs 
portals is much less constrained than in the standard picture, see Fig.~\ref{plots_omega_current} for the current exclusion limits. In particular, none of the existing bounds limits the region of masses $m_S>\unit[200]{GeV}$ for couplings $\lambda_S$ smaller than 1, except for small regions of the parameter space. Only a limited band within this region  will be probed by the next generation of direct detection experiments, see Figs.~\ref{plot_summaryc101} and ~\ref{plot_summaryc201} for XENON1T~\cite{Aprile:2012zx} prospects.

The viable parameter spaces differ so much between the two scenarios, that it may be possible to single out signals excluding the standard portal. Let us suppose, for example, that Xenon1T observes a DM 
signal at a mass $m_S\simeq \unit[200]{GeV}$, measuring a DM-nucleon scattering 
cross-section with some value $\hat{\s}_{SI}$. In the standard Higgs-portal 
interpretation, this would give a point in the $(m_S,\lambda_S)$ plane: the 
coupling is uniquely determined by the values of the mass and of the 
cross-section. In a non-linear portal setting, instead, the measure would 
translate into a viable vertical line whose size depends on the values assumed 
for the non-linear coefficients.  Now, it may happen that the point in the linear plane 
falls within a region which is already ruled out (for example by Planck or by some 
collider constraint), while the line in the non-linear plot is (at least 
partly) allowed. This kind of signals would represent a strong indication in favour of 
extra interactions beyond the standard Higgs portal.

Another characteristic aspect of non-linear portals is the enhancement of signal rates at colliders. In this paper we studied production of a pair of DM particles in association with a vector boson or a Higgs. In the standard Higgs portal, the production of DM particles is unique: a Higgs produced in gluon fusion radiating two DM particles. This production is very suppressed for DM heavier or around the Higgs mass, whereas light DM appears already excluded by a combination of Higgs invisible width, relic abundance and  direct detection constraints. 
Non-linear interactions allow electroweak production of DM via couplings to vector bosons, leading to mono-$W$, mono-$Z$ and mono-Higgs signatures with rates ${\cal O}(10^{1-4}) \times c_i^2$ bigger than the standard Higgs.  Additionally, these new production modes exhibit specific kinematic features which may help in disentangling standard and non-linear production. We have shown that a smoking gun to distinguish the standard portal from the non-linear one is provided by the combined measurement of the cross-sections ratio $R_{WZ}=\s(pp\to ZSS)/\s(pp\to W^\pm SS)$ with that of $m_S$ from transverse momentum distributions.

For comparative purposes between the linear and non-linear expansions, as part of the theoretical analysis we have determined  the linear siblings of all couplings studied. We determined the complete basis of purely bosonic  $d=6$ operators of the linear realisation  and also the subset of linear $d=8$ operators  which induce the same physical couplings as those in the non-linear portal,  up to two Dark Matter fields.  While all operators of the non-linear portal considered appear at leading order, their siblings are subleading corrections in the linear expansion and their amplitude should be duly suppressed. Nevertheless, we have discussed how to distinguish the impact of both expansions, in case the relative amplitude of a $d\ge6$ linear operator  becomes enhanced due to some fine-tuning. A tool to disentangle the impact of higher-dimension linear operators from the leading non-linear ones may result, in principle, from the analysis of (de)correlations  of specific couplings: $S-S-Z$ {\it vs.} $S-S-Z-h$ and $S-S-Z-Z$ {\it vs.} $S-S-W-W$.  
 Finally, note that the features and bounds obtained in the analysis of the non-linear portal apply equally well to the standard one, except in regions of the parameter space which undergo restrictions due to constraints on the cut-off of the theory. 

The search for Dark Matter and the quest for the nature of electroweak symmetry breaking are major present challenges. We have discussed their interplay within an effective approach, in the framework of the Higgs Dark Matter portal.
 

\section*{Acknowledgements}
We  thank A. Manohar and E. Jenkins for useful discussions and for reading  the manuscript. 
The work of K.M. and V.S. is supported by the Science Technology and Facilities Council (STFC) under grant number
ST/L000504/1. 
I.B. research was supported by an ESR contract of the EU network FP7 ITN INVISIBLES (Marie Curie Actions, PITN-GA-2011-289442). I.B., M.B.G., L.M., R.dR. acknowledge partial 
support of the European Union network FP7 ITN INVISIBLES, of CiCYT through the project FPA2012-31880 and of the Spanish MINECO's ``Centro de Excelencia Severo Ochoa'' Programme under 
grant SEV-2012-0249. M.B.G. and L.M. acknowledge partial support by a grant from the Simons Foundation and  the Aspen Center for Physics, where part of this work has been developed, 
which is supported by National Science Foundation grant PHY-1066293.
J.M.N. is supported by the People Programme (Marie Curie Actions) of the European Union Seventh Framework Programme (FP7/2007-2013) under REA grant agreement PIEF-GA-2013-625809.

\clearpage
\appendix

\section{Feynman rules}
\label{Feynman_rules}
This Appendix provides a complete list of the Feynman rules resulting from the non-linear Higgs portal effective Lagrangian, Eq.~(\ref{L0S}), 
computed in unitary gauge and with momenta understood to flow inwards. The right column shows for comparison the Feynman rules for the case 
of the linear Higgs portal $\lambda_{S}\,S^2\,(2vh + h^2)$.

\newcommand{\nr}{\stepcounter{diagram}(FR.\arabic{diagram})}
\newcounter{diagram}
\renewcommand{\arraystretch}{6}
\newlength{\FDwidth}
\setlength{\FDwidth}{2.5cm}

\vspace*{1cm}
\hspace*{-5mm}\noindent
{\footnotesize\begin{tabular}
{l@{\hspace*{5mm}}>
{\centering}p{\FDwidth}>
{\centering$\displaystyle}m{1.7cm}<{$}>
{\centering$\displaystyle}m{4.5cm}<{$}>
{\centering$\displaystyle}m{4.5cm}<{$}}
 \toprule
 \\[-3.7cm]
 & 
 & \text{\bf Standard }
 & \text{\bf Non-linear}
 & \text{\bf Linear $d\leq6$}
 \tabularnewline[-3mm]\midrule
 
\nr&\parbox{\FDwidth}{
\begin{fmffile}{Fdiagrams/SSHcoupl}
\begin{fmfgraph*}(50,50)

\fmfright{i1,i2,i3}
\fmfleft{o3,o2,o1}
  \fmf{dashes}{i2,v1}
  \fmf{dashes}{v1,o1}
  \fmf{dashes}{v1,o3}
  
\fmfv{lab=$S$,l.angle=0}{o1}
\fmfv{lab=$S$,l.angle=0}{o3}
\fmfv{lab=$h$}{i2}
\end{fmfgraph*}
\end{fmffile}
} 
&-4i \lambda_{S} v
& -4i\left(\lambda_{S} v+\frac{c_2a_2p_h^2}{v}\right)
& -4i\left(\lambda_{S} v+\frac{2vc^L_2p_h^2}{\Lambda^2}\right)
\tabularnewline

\nr& \parbox{\FDwidth}{
\begin{fmffile}{Fdiagrams/SSZcoupl}
\begin{fmfgraph*}(50,50)

\fmfright{i1,i2,i3}
\fmfleft{o3,o2,o1}
  \fmf{boson}{i2,v1}
  \fmf{dashes}{v1,o1}
  \fmf{dashes}{v1,o3}
  
\fmfv{lab=$S$,l.angle=0}{o1}
\fmfv{lab=$S$,l.angle=0}{o3}
\fmfv{lab=$Z_\mu$}{i2}
\end{fmfgraph*}
\end{fmffile}
} 
& -
& \frac{2gc_4}{\ct}p_{Z}^\mu
& -\frac{4v^2gc^L_4}{\ct\Lambda^2}p_{Z}^\mu
\tabularnewline
  
\nr& \parbox{\FDwidth}{\centering
\begin{fmffile}{Fdiagrams/SSHHcoupl}
\begin{fmfgraph*}(50,50)

\fmfleft{i1,i2}
\fmfright{o2,o1}
  \fmf{dashes}{i2,v1}
  \fmf{dashes}{i1,v1}
  \fmf{dashes}{v1,o1}
  \fmf{dashes}{v1,o2}
  
\fmfv{lab=$h$,l.angle=0}{o1}
\fmfv{lab=$h$,l.angle=0}{o2}
\fmfv{lab=$S$,l.angle=180}{i1}
\fmfv{lab=$S$,l.angle=180}{i2}
\end{fmfgraph*}
\end{fmffile}
} 
&-4i\lambda_{S}
& -4i\left(\lambda_{S} b +\frac{c_2b_2 (p_{h1}+p_{h2})^2}{v^2}\right) 
& -4i\left(\lambda_{S} +\frac{3v^2c_b}{2\Lambda^2}+\frac{2c^L_2 (p_{h1}+p_{h2})^2}{\Lambda^2}\right) 
\tabularnewline

\nr& \parbox{\FDwidth}{\centering
\begin{fmffile}{Fdiagrams/SSZZcoupl}
\begin{fmfgraph*}(50,50)

\fmfleft{i1,i2}
\fmfright{o2,o1}
  \fmf{dashes}{i2,v1}
  \fmf{dashes}{i1,v1}
  \fmf{boson}{v1,o1}
  \fmf{boson}{v1,o2}
  
\fmfv{lab=$Z_\mu$,l.angle=0}{o1}
\fmfv{lab=$Z_\nu$,l.angle=0}{o2}
\fmfv{lab=$S$,l.angle=180}{i1}
\fmfv{lab=$S$,l.angle=180}{i2}
\end{fmfgraph*}
\end{fmffile}
} 
& -
&-\frac{2ig^2(c_1+2c_3)}{\ct^2}g_{\mu\nu}
&-\frac{8v^2ig^2c^L_1}{\ct^2\Lambda^2}g_{\mu\nu}
\tabularnewline

\nr& \parbox{\FDwidth}{\centering
\begin{fmffile}{Fdiagrams/SSWWcoupl}
\begin{fmfgraph*}(50,50)

\fmfleft{i1,i2}
\fmfright{o2,o1}
  \fmf{dashes}{i2,v1}
  \fmf{dashes}{i1,v1}
  \fmf{boson}{v1,o1}
  \fmf{boson}{v1,o2}
  
\fmfv{lab=$W^+_\mu$,l.angle=0}{o1}
\fmfv{lab=$W^-_\nu$,l.angle=0}{o2}
\fmfv{lab=$S$,l.angle=180}{i1}
\fmfv{lab=$S$,l.angle=180}{i2}
\end{fmfgraph*}
\end{fmffile}
} 
& -
& -2ig^2 c_1 g_{\mu\nu}
& -8\frac{v^2}{\Lambda^2}ig^2 c^L_1 g_{\mu\nu}
\tabularnewline

\nr& \parbox{\FDwidth}{\centering
\begin{fmffile}{Fdiagrams/SSZHcoupl}
\begin{fmfgraph*}(50,50)

\fmfleft{i1,i2}
\fmfright{o2,o1}
  \fmf{dashes}{i2,v1}
  \fmf{dashes}{i1,v1}
  \fmf{boson}{v1,o1}
  \fmf{dashes}{v1,o2}
  
\fmfv{lab=$Z_\mu$,l.angle=0}{o1}
\fmfv{lab=$h$,l.angle=0}{o2}
\fmfv{lab=$S$,l.angle=180}{i1}
\fmfv{lab=$S$,l.angle=180}{i2}
\end{fmfgraph*}
\end{fmffile}
} 
& -
& \frac{4g}{v\ct}\left(c_4a_4 (p_Z+p_h)^\mu-c_5 a_5 p_h^\mu\right)
& -\frac{8vg}{\Lambda^2\ct}\left(c^L_4 (p_Z+p_h)^\mu\right)
\tabularnewline
\bottomrule
\end{tabular}}

\newpage

\section{Contributions to the Dark Matter relic abundance}
\label{Relic_Abundance}

The Feynman diagrams contributing to the 
main Higgs portal DM annihilation processes are shown next. The labels indicate the parameters entering each vertex (see 
Appendix \ref{Feynman_rules} for signs and numerical factors). $\lambda_h$ in \ref{DM_hh} stands for the SM Higgs self-coupling.

\begin{figure}[ht!]\centering
\begin{subfigure}{\textwidth}\centering
\input{Fdiagrams/SSHH_annihilation}
\caption{Dark Matter annihilation to Higgs bosons.}\label{DM_hh}
\end{subfigure}
\begin{subfigure}{\textwidth}\centering
\input{Fdiagrams/SSWW_annihilation}
\caption{Dark Matter annihilation to $W$ bosons.}\label{DM_WW}
\end{subfigure}
\begin{subfigure}{\textwidth}\centering
\input{Fdiagrams/SSZZ_annihilation}
\caption{Dark Matter annihilation to $Z$ bosons.}\label{DM_ZZ}
\end{subfigure}
\begin{subfigure}{\textwidth}\centering
\input{Fdiagrams/SSZh_annihilation}
\caption{Dark Matter annihilation to $Z$ and Higgs bosons.}\label{DM_Zh}
\end{subfigure}
\begin{subfigure}{\textwidth}\centering
\input{Fdiagrams/SSbb_annihilation}
\caption{Dark Matter annihilation to $f\bar{f}$.}\label{DM_bb}
\end{subfigure}
\end{figure}

\clearpage
\section{Impact of \texorpdfstring{$\A_1$ and $\A_2$}{A1} for other choices of \texorpdfstring{$c_i$}{ci}}\label{c12_extras}
The analysis of the current constraints on the parameter space of non-linear Higgs portals described in Section~\ref{Sect:Pheno} is restricted to two specific 
non-linear setups: fixing either $c_1$ or $c_2$ to 0.1 (see Figure~\ref{plots_summarynonlin}).
Although the main features of non-linearity are quite exhaustively illustrated by these two examples, it is interesting to explore further scenarios, 
where the coefficients $c_1$ and $c_2$ are assigned different values in the range $[-1,1]$. In this Appendix we show the exclusion 
regions obtained for $c_i=\{\pm1,-0.1,-0.01\}$ and $c_2=\pm1$. These figures shall be compared with Figure~\ref{plots_summarylin}, where the same constraints 
have been applied to the linear Higgs-portal scenario.

\vspace{2mm}

\begin{figure}[t!]\centering
\begin{subfigure}{.495\textwidth}\centering
\hspace*{-1cm}
\includegraphics[height=6.6cm]{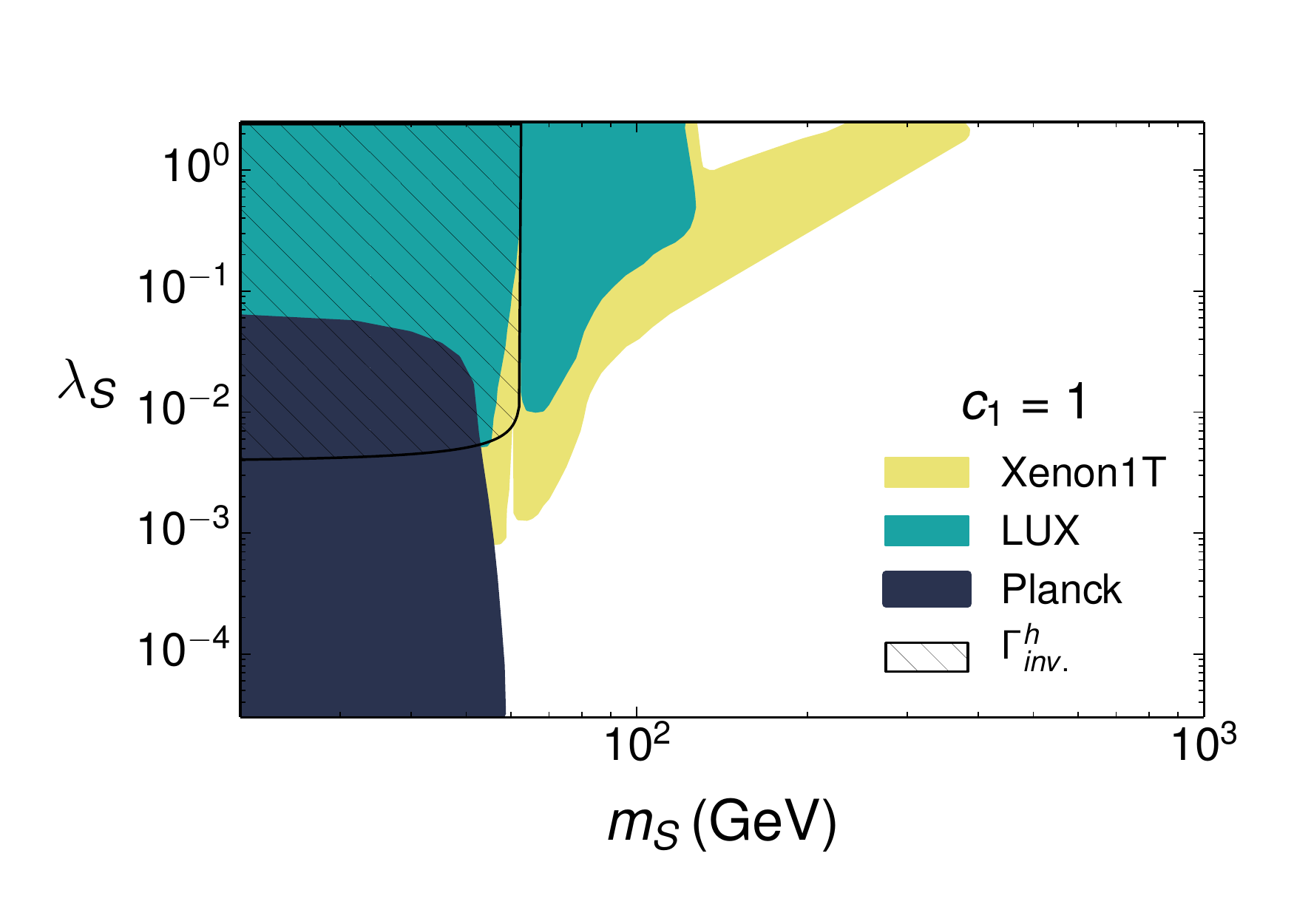}
\caption{$c_1=1$}\label{plot_summaryc11}
\end{subfigure}
\hfill
\begin{subfigure}{.495\textwidth}\centering
\includegraphics[trim = 1.3cm 0cm 0cm 0cm, clip=true, totalheight=6.6cm]{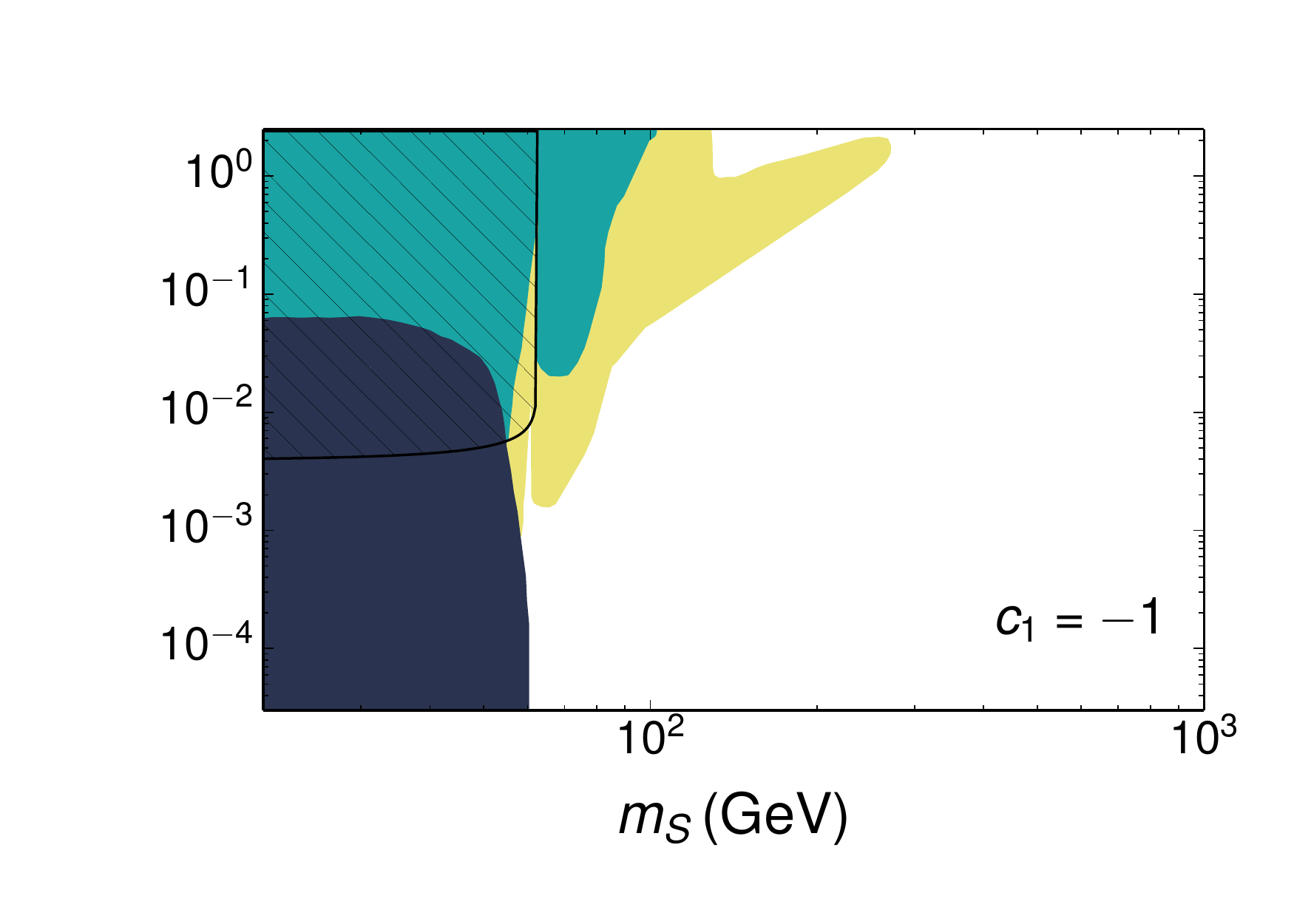}
\caption{$c_1=-1$}\label{plot_summaryc1m1}
\end{subfigure}\\
\begin{subfigure}{.495\textwidth}\centering
\hspace*{-1cm}
\includegraphics[height=6.6cm]{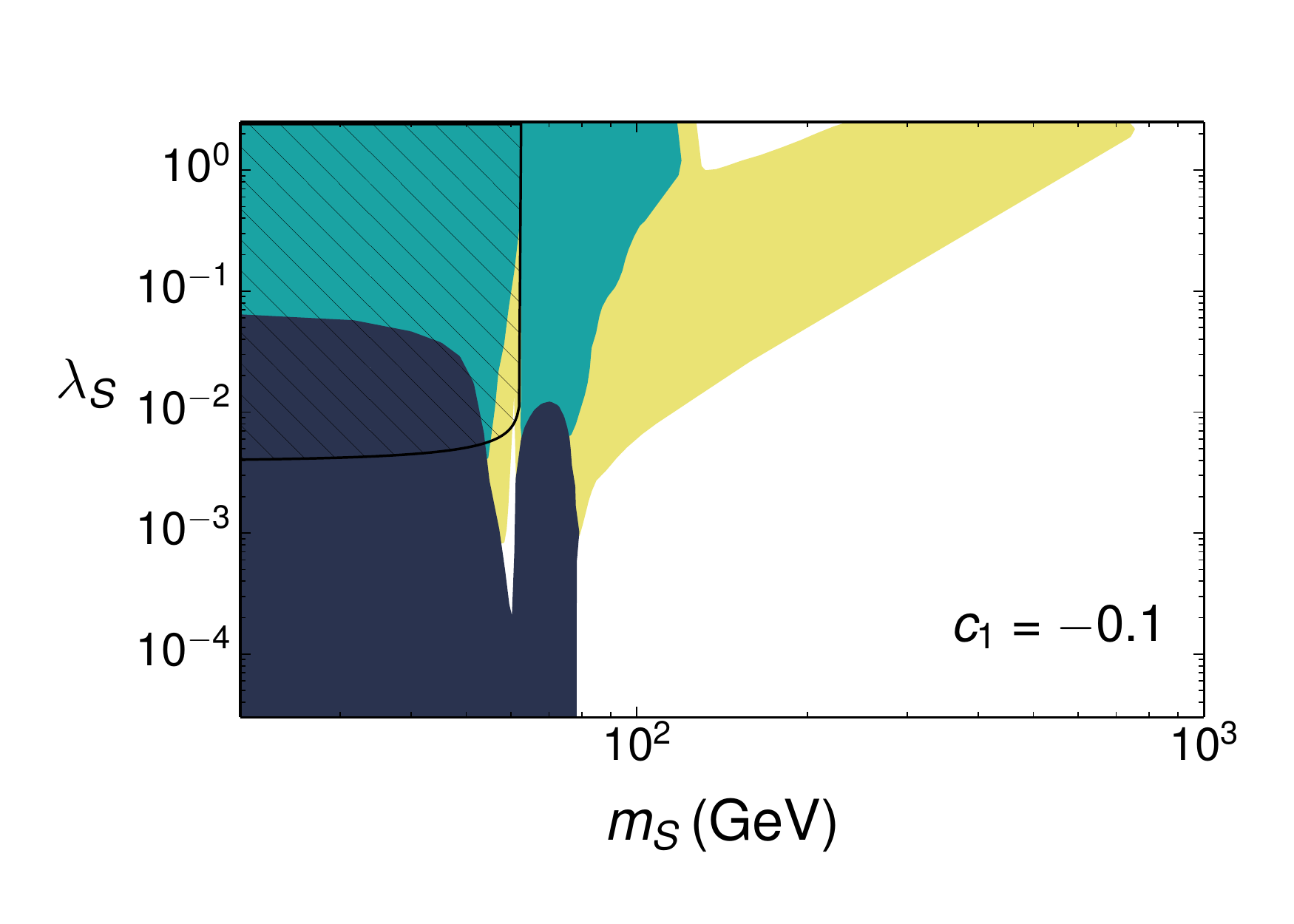}
\caption{$c_1=-0.1$}\label{plot_summaryc1m01}
\end{subfigure}
 \hfill
\begin{subfigure}{.495\textwidth}\centering
\includegraphics[trim = 1.3cm 0cm 0cm 0cm, clip=true, totalheight=6.6cm]{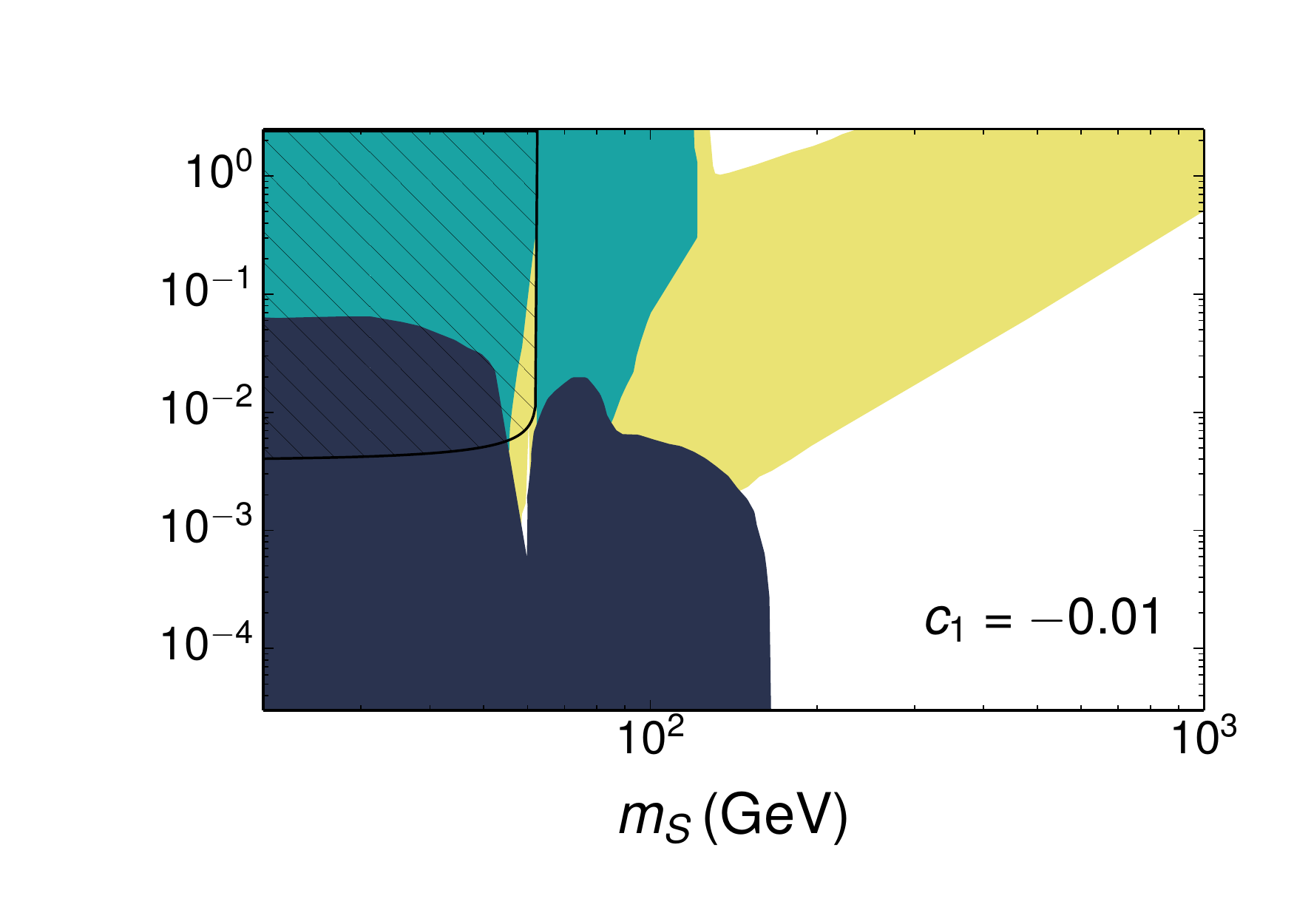}
\caption{\it $c_1=-0.01$}\label{plot_summaryc1m001}
\end{subfigure}
\caption{Results obtained considering the non-linear operator $\A_1$ with $\F_1(h)=(1+h/v)^2$ and for different values of 
the coefficient $c_1$. The blue region is excluded by current bounds from 
Planck, the green one is excluded by LUX, while the area in yellow is within 
the projected reach of XENON1T. The black hatched region represents the bound from 
invisible Higgs width (same as in the linear scenario).}\label{plots_summaryc1} 
\end{figure}

As a general feature, it is worth noticing that in presence of non-linearity, even conveyed by a coefficient of order $0.1$ 
(Figures~\ref{plots_summarynonlin} and~\ref{plot_summaryc1m01}) or $0.01$ (Figure~\ref{plot_summaryc1m001}), the space of parameters for Higgs 
portals is much less constrained than in the standard picture. In particular, none of the existing bounds limit the region of masses $m_S>\unit[200]{GeV}$ 
for couplings $\lambda_S$ smaller than 1, except for small regions of the parameter space. Only a limited band within this region will be probed by the next generation of direct detection experiments (the 
plots show the reach of XENON1T~\cite{Aprile:2012zx}).

\begin{figure}[t!]\centering
\begin{subfigure}{.495\textwidth}\centering
\hspace*{-1cm}
\includegraphics[height=6.6cm]{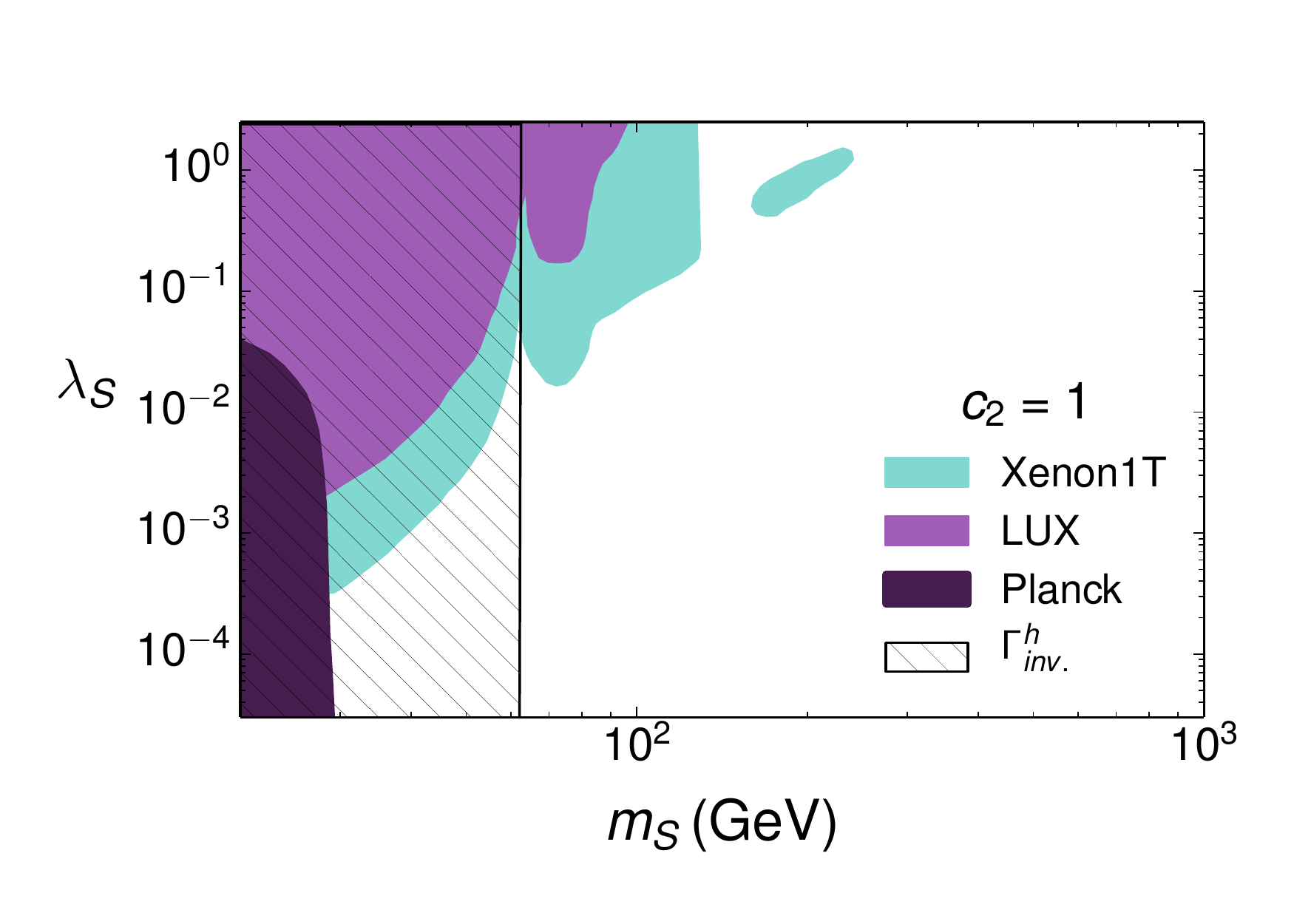}
\caption{$c_2=1$}\label{plot_summaryc211}
\end{subfigure}
\hfill
\begin{subfigure}{.495\textwidth}\centering
\includegraphics[trim = 1.3cm 0cm 0cm 0cm, clip=true, totalheight=6.6cm]{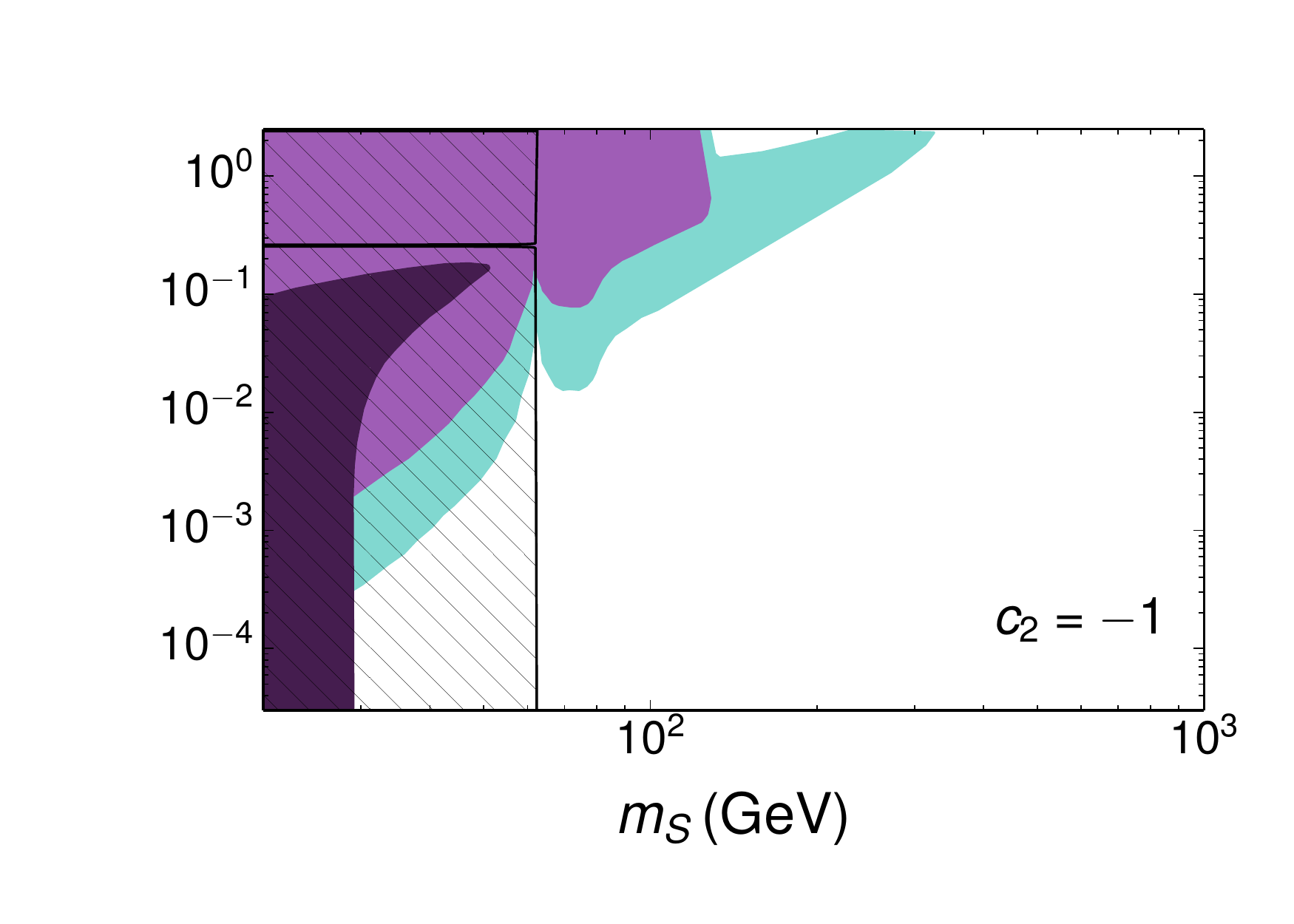}
\caption{$c_2=-1$}\label{plot_summaryc2m1}
\end{subfigure}
\caption{Results obtained considering the 
non-linear operator $\A_2$ with $\F_2(h)=(1+h/v)^2$ and for $c_2=\pm1$. The darkest region is excluded by current bounds from 
Planck, the purple one is excluded by LUX, while the area in light blue is within 
the projected reach of XENON1T.  The black hatched region represents the bound 
from invisible Higgs width.}
\label{plots_summaryc2} 
\end{figure}


\clearpage
\providecommand{\href}[2]{#2}\begingroup\raggedright\endgroup

\end{document}